\documentclass[twoside,twocolumn,english,prl]{revtex4-1}
\usepackage[T1]{fontenc}
\usepackage[latin9]{inputenc}
\usepackage{geometry}
\usepackage{graphicx}

\makeatletter
\@ifundefined{date}{}{\date{}}
\usepackage{babel}

\makeatother

\usepackage{babel}
\usepackage{times}
\usepackage{graphicx}
\usepackage{amssymb}
\usepackage{epstopdf}

\begin{document}

\title{Magnetic anisotropy of the alkali iridate Na$_{2}$IrO$_{3}$ at high magnetic fields: evidence for strong ferromagnetic Kitaev correlations}

\author{Sitikantha D. Das$^{1,10}$}
\author{Sarbajaya Kundu$^{2}$}
\author{Zengwei Zhu$^{3}$}
\altaffiliation[Now at: ]{National High Magnetic Field Center and School of Physics, Huazhong University of Science and Technology, Wuhan 430074, China}
\author{Eundeok Mun$^{3}$}
\altaffiliation[Now at: ]{Department of Physics, Simon Fraser University, Burnaby, BC, Canada V5A-1S6}
\author{Ross D. McDonald$^{3}$}
\author{Gang Li$^{4}$}
\altaffiliation[Now at: ]{Institute of Physics, Chinese Academy of Sciences, P.O Box 603, Beijing 100190, China}
\author{Luis Balicas$^{4}$}
\author{Alix McCollam$^{5}$}
\author{Gang Cao$^{6,7}$}
\author{Jeffrey G. Rau$^{8}$}
\author{Hae-Young Kee$^{8,9}$}
\author{Vikram Tripathi$^{2}$}
\author{Suchitra E. Sebastian$^{1}$}

\affiliation{$^{1}$Cavendish Laboratory, University of Cambridge, J J Thomson Avenue, Cambridge CB3 0HE, UK}
\affiliation{$^{2}$Department of Theoretical Physics, Tata Institute of Fundamental Research, Homi Bhabha Road, Colaba, Mumbai 400005, India}
\affiliation{$^{3}$Los Alamos National Laboratory, Los Alamos, New Mexico 87545, USA}
\affiliation{$^{4}$National High Magnetic Field Laboratory,1800 E. Paul Dirac Drive, Tallahassee, FL 32310, USA}
\affiliation{$^{5}$High Field Magnet Laboratory (HFML - EMFL), Radboud University, 6525 ED, Nijmegen, The Netherlands}
\affiliation{$^{6}$Center for Advanced Materials and Department of Physics and Astronomy, University of Kentucky, Lexington, Kentucky 40506, USA}
\affiliation{$^{7}$Department of Physics, 390 UCB, University of Colorado, Boulder, CO 80309, USA}
\affiliation{$^{8}$Department of Physics, University of Toronto, Toronto, Ontario M5S 1A7, Canada}
\affiliation{$^{9}$Canadian Institute for Advanced Research/Quantum Materials Program, Toronto, Ontario MSG 1Z8, Canada}
\affiliation{$^{10}$Department of Physics, IIT, Kharagpur, Kharagpur 721302, India}

\date{\today}
\begin{abstract}
The magnetic field response of the Mott-insulating honeycomb iridate Na$_{2}$IrO$_{3}$ is investigated using torque magnetometry measurements in magnetic fields up to 60 tesla.
A peak-dip structure is observed in the torque response at magnetic fields corresponding to an energy scale close to the zigzag ordering ($\approx 15$K) temperature.
Using exact diagonalization calculations, we show that such a distinctive signature in the torque response 
constrains the effective spin models for these classes of Kitaev materials to ones with dominant ferromagnetic Kitaev interactions, while alternative models with dominant 
antiferromagnetic Kitaev interactions are excluded. We further show that at high magnetic fields, long range spin 
correlation functions decay rapidly, signaling a transition to a long-sought-after field-induced quantum spin liquid beyond the peak-dip structure.
Kitaev systems are thus revealed to be excellent candidates for field-induced quantum spin liquids, similar physics having been suggested in another Kitaev material $\alpha-$RuCl$_{3}$.
\end{abstract}
\maketitle
The alkali iridates A$_{2}$IrO$_{3}$(A=Na,Li), along with their celebrated 4d analogue, $\alpha-$RuCl$_{3}$
~\cite{PhysRevLett.118.107203,PhysRevB.93.075144,PhysRevLett.114.147201,PhysRevB.90.041112,PhysRevB.91.241110,PhysRevB.93.155143,PhysRevB.93.134423,PhysRevB.94.020407,PhysRevB.96.054410,
PhysRevB.91.180401,winter2017breakdown,Banerjee2016proximate}, 
have attracted much theoretical \cite{chaloupka2010kitaev,kimchi2011kitaev,katukuri2014kitaev,
sizyuk2014importance,rau2014trigonal,rau2015spin,yamaji2014first,PhysRevB.92.024413,PhysRevB.84.100406,PhysRevB.98.094401,
yao2015zigzag,bhattacharjee2012spin,hu2015first,foyevtsova2013ab,jiang2011possible,chaloupka2013zigzag}
and experimental \cite{singh2012relevance,ye2012direct,choi2012spin,singh2010antiferromagnetic,chun2015direct,comin20122,clancy2012spin,gretarsson2013crystal,gretarsson2013magnetic,liu2011long,
Banerjee2018excitations,mehlawat2017heat}
attention as promising candidates for realizing the physics of the honeycomb Kitaev model \cite{kitaev2003fault,kitaev2006anyons}.  Interactions between the effective $j_{\rm eff}=\frac{1}{2}$ 
pseudospins on every site of the two-dimensional hexagonal lattice in these strongly spin-orbit coupled materials, have been described by a dominant Kitaev and other subdominant interactions 
such as Heisenberg \cite{jackeli2009mott} and symmetric off-diagonal exchange \cite{kim2008novel,gretarsson2013crystal,jackeli2009mott,singh2010antiferromagnetic,rau2014trigonal,rau2014generic}. 
Notwithstanding the great progress made, the sign of the dominant Kitaev interaction, vital for ascertaining the correct physics of these materials
~\cite{katukuri2014kitaev,sizyuk2014importance,chaloupka2013zigzag,Cookmeyer2018spin,koitzsch2017low}, remains an open question. The importance of the 
magnetic field response in determining the same has been emphasized in multiple studies recently \cite{yadav2016kitaev,janssen2017magnetization}, and it has been used to experimentally 
investigate the Kitaev material $\alpha-$RuCl$_{3}$~\cite{leahy2017anomalous}. Yet high field studies have thus far been impracticable in $\rm{Na}_{2}$IrO$_{3}$ because of the evidently higher 
energy scales involved. Here, we probe the physics of $\rm{Na}_{2}$IrO$_{3}$ by using a combination of magnetometry studies at high magnetic fields up to 60~T, and exact diagonalization calculations. 
We find a distinctive peak-dip structure in the experimental torque response at high fields, which we use to constrain the model description of $\rm{Na}_{2}$IrO$_{3}$. 
By comparison with results of exact diagonalisation calculations, we show that this nonmonotonic signature is uniquely captured by a model with a 
dominant ferromagnetic Kitaev exchange ~\cite{katukuri2014kitaev,kimchi2011kitaev,sizyuk2014importance,singh2012relevance}, 
but not one with an antiferromagnetic Kitaev ~\cite{chaloupka2013zigzag,rau2014generic} counterpart. We also find that the finely-tuned zigzag ground state, expected for such a model, 
gives way to a quantum spin liquid 
state by field tuning beyond the peak-dip feature. 
Intriguingly, a similar feature in the anisotropic magnetisation
has also been observed in $\alpha-$RuCl$_{3}$, but not explained ~\cite{leahy2017anomalous,Riedl2018on}.
Here we show the likely universality of such a feature in the magnetic torque,
as a signature of the field-induced spin liquid (also revealed in the Kitaev system  $\alpha-$RuCl$_{3}$  at lower energy scales \cite{PhysRevLett.119.227208,PhysRevB.98.094414,PhysRevB.95.241112}),
thus shedding light on the relevance of Kitaev materials for realising a quantum spin liquid ground state.

Na$_{2}$IrO$_{3}$ is a layered Mott insulator with an energy gap $E_{g}$ = 340 meV\cite{comin20122} and spin-orbit coupling $\lambda\approx 0.5$ eV\cite{rau2015spin}. The 
material is highly frustated magnetically, with a Curie-Weiss temperature of $\theta_{\rm{CW}}\approx -116$ K and a  N$\acute{e}e$l temperature of $T_{N} \approx 15$ K. 
\cite{singh2010antiferromagnetic,chun2015direct,ye2012direct,choi2012spin}. From Neutron and X-ray diffraction\cite{ye2012direct}, inelastic neutron scattering(INS)\cite{choi2012spin} 
and resonant inelastic X-ray scattering(RIXS)\cite{liu2011long} measurements, the ground state is known to be an antiferromagnetic zigzag phase with 
an ordered moment $\mu_{\rm{ord}}\approx 0.2\mu_{B}$\cite{singh2010antiferromagnetic,ye2012direct,choi2012spin}. 
The parameter space for couplings in Na$_{2}$IrO$_{3}$ has thus far been constrained using ab-initio computations\cite{katukuri2014kitaev,yamaji2014first,hu2015first,foyevtsova2013ab}, 
 exact diagonalization\cite{chaloupka2013zigzag,rau2014trigonal,rau2014generic,PhysRevB.94.064435}\textbf{}, classical \textbf{ }Monte Carlo simulations
\cite{sizyuk2014importance,yao2015zigzag}, 
degenerate perturbation theory\cite{chaloupka2010kitaev,chaloupka2013zigzag,kimchi2011kitaev,rau2014trigonal,rau2014generic}, as well as experimental investigation\cite{choi2012spin}. 
The simplest model arrived at is a nearest-neighbor model with a dominant \emph{antiferromagnetic} Kitaev~\cite{chaloupka2013zigzag,rau2014generic} and a smaller 
\emph{ferromagnetic} Heisenberg exchange. In subsequent calculations we refer to this model as \emph{Model A}. 
A different model with a dominant \emph{ferromagnetic} 
Kitaev and smaller \emph{antiferromagnetic} Heisenberg exchange is however suggested by quantum chemistry~\cite{katukuri2014kitaev} and other ab-initio calculations~\cite{yamaji2014first,
hu2015first,foyevtsova2013ab}. 
In order to stabilize a zigzag phase within such a model, we consider variants of this model with further neighbor
couplings\cite{foyevtsova2013ab,sizyuk2014importance,katukuri2014kitaev,kimchi2011kitaev,singh2012relevance} 
 (\emph{Model B}), or additional anisotropic interactions\cite{yamaji2014first} (\emph{Model C}). 
Here we distinguish between 
models with either dominant antiferromagnetic or ferromagnetic Kitaev interactions, by measuring the finite-field response of Na$_{2}$IrO$_{3}$ and comparing our results
with exact diagonalization simulations.

\begin{figure}
\includegraphics[width=0.7\columnwidth]{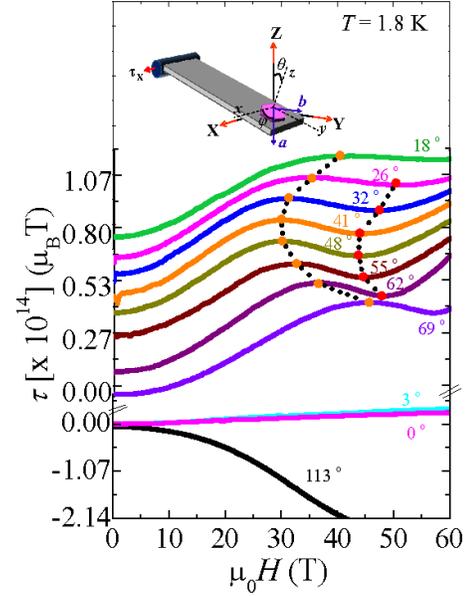}

\caption{\label{fig:tau-1} 
Measure torque ($\tau$) as a function of magnetic field for different polar angles ($\theta$) and azimuthal angle $\phi = 90^{\circ}$. A peak dip structure is observed in the torque, 
and is seen to evolve with $\theta$. Individual torque curves have been offset for clarity. (inset: a crystal on the cantilever with the
various coordinate systems: $XYZ\rightarrow$lab frame, $xyz\rightarrow$frame
fixed to the cantilever, so that $X$ and $x$ coincide. $\theta$
is the angle that the normal to the crystal makes with the magnetic
field, and the measured magnetic torque along the $X$-direction is referred to as $\tau$.)}
\end{figure}

\begin{figure}
\includegraphics[width=0.7\columnwidth]{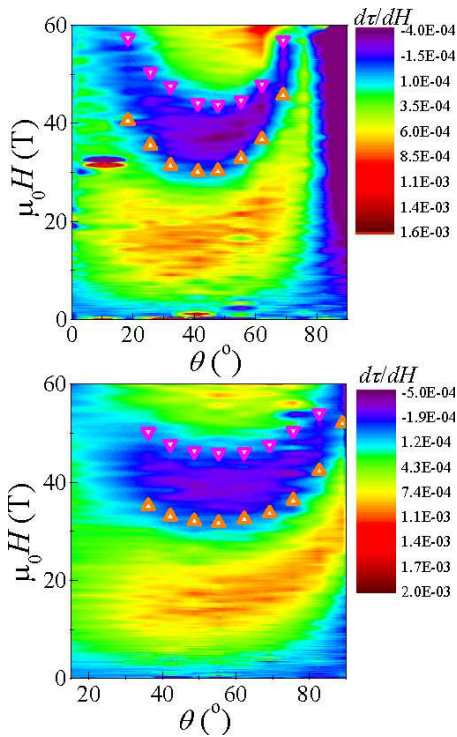}

\caption{\label{fig:cheshirecat-1} Derivative of experimentally measured magnetic torque with respect to field \big($\frac{d\tau}{dH}$\big) as a function of 
field and angle ($\theta$) for $\phi=90^{\circ}$ (Top) and $\phi=0^{\circ}$
(Bottom) . The position of the maxima in the torque is indicated by regular triangles while that of the subsequent minima is marked by inverted triangles. }
\end{figure}

A single crystal of Na$_2$IrO$_3$, of dimension $\approx$~100 $\mu$m on a side, with a much smaller thickness, was mounted on a piezoresistive cantilever and measured on an in-situ rotating stage in 
pulsed magnetic fields up to 60~T.
The torque response($\tau$) was measured as a function of the magnetic field at various fixed angles ($0^{\circ}\lesssim\theta\lesssim 90^{\circ}$) of the crystalline axis normal to 
the honeycomb lattice, with respect to the magnetic field axis. A distinctive non-monotonic feature is observed in the torque response~(Fig.~\ref{fig:tau-1}). 
A peak in the magnetic torque in the vicinity of 30-40 T 
is followed by a dip in the vicinity of 45-55 T. The peak and dip features are separated by as much as $\approx$15~T near $\theta \approx 45^\circ-55^\circ$, but draw closer
together at angles closer to $\theta \approx 0^\circ$ and $\theta \approx 90^\circ$. In the vicinity of $\theta \approx 0^\circ$ and $\theta \approx 90^\circ$, the peak and dip features
are seen to merge into a single plateau-like feature. This evolution of the signature peak-dip feature as a function of field-inclination angle and magnetic field is shown in Fig.~\ref{fig:cheshirecat-1} 
for two different azimuthal orientations ($\phi=0^{\circ},90^{\circ}$), where $\phi$ is the angle that the crystallographic $a$ axis makes with the axis of
rotation of the cantilever. The high magnetic field torque response of Na$_{2}$IrO$_{3}$ was independently measured for two crystals, for three different azimuthal 
orientations ($\phi=0^{\circ},90^{\circ}$and $180^{\circ}$), at a temperature of 1.8 K and results for both were found to be very similar (data for the second sample is shown in the SI).
The signature peak-dip feature is found to disappear above the zigzag ordering temperature (SI). Meanwhile, the isotropic magnetization($m_{Z}$) measured using an extraction magnetometer
in pulsed magnetic fields up to 60 T, and a force magnetometer in steady fields up to 30~T~\cite{mccollam1}, increases linearly with
magnetic field up to 60~T (SI).

We use theoretical modeling of the non-monotonic features in the high field response to distinguish between potential microscopic models.
Our starting point is the usual spin Hamiltonian\cite{chaloupka2010kitaev,chaloupka2013zigzag} with nearest-neighbor Kitaev and Heisenberg interactions: 
\begin{equation}
J_{\rm{h}}\sum_{<ij>}\overrightarrow{\sigma_{i}}.\overrightarrow{\sigma_{j}}+J_{\rm{K}}\sum_{<ij>}\sigma_{i}^{\gamma}\sigma_{j}^{\gamma}\label{eq:2-1}
\end{equation}
where $\gamma=x,y,z$ labels an axis in spin space and a bond direction of the honeycomb lattice. 
\emph{Model A} is parametrised by nearest-neighbour interactions $J_{\rm{h}}<0$ and  $J_{\rm{K}}>0$. In \emph{Model B}, further neighbor 
antiferromagnetic Heisenberg couplings $J_{2}$ and $J_{3}$ \cite{kimchi2011kitaev} are introduced up to the third nearest neighbor,
 with $J_{\rm{h}}>0$ and  $J_{\rm{K}}<0$. In \emph{Model C}, bond-dependent nearest-neighbor symmetric off-diagonal 
 terms $H_{\rm{od}}^{(\gamma)}=\Gamma\sum_{\alpha\neq\beta\neq\gamma}\sum_{\{i,j\}}(\sigma_{i}^{\alpha}\sigma_{j}^{\beta}+\sigma_{i}^
{\beta}\sigma_{j}^{\alpha})$ 
(where $\alpha$ and $\beta$ are the two remaining directions apart from the Kitaev bond direction $\gamma$) \cite{rau2014generic} and
$H_{\rm{od}}^{\prime}=\Gamma^{\prime}\sum_{\alpha\neq\beta\neq\gamma}\sum_{\{i,j\}}(\sigma_{i}^{\beta}\sigma_{j}^{\gamma}+\sigma_{i}^{\gamma}\sigma_{j}^{\beta}+\sigma_{i}^{\alpha}\sigma_{j}^{\gamma}+\sigma_
{i}^{\gamma}\sigma_{j}^{\alpha})$
\cite{rau2014trigonal} accounting for trigonal distortions of the oxygen octahedra, are introduced. The main features of these models 
are summarized in Table I. 

For our calculations, we use a hexagonal 24-site cluster \cite{chaloupka2013zigzag,chaloupka2010kitaev,rau2014generic,rau2014trigonal} 
with periodic boundary conditions. The effect of the applied field $\overrightarrow{H}=H\hat{z}$ (in the lab frame) on the system is described by
$H_{\rm{mag}}=(\frac{g}{2})\sum_{i}\sum_{\gamma}h_{\gamma}\sigma_{i}^{\gamma}$, with $g\approx 1.78$ \cite{chaloupka2013zigzag} and 
$\overrightarrow{h}=(h_{x},h_{y},h_{z})$ being the field as expressed in the crystal octahedron frame. Exact diagonalization calculations for the ground state energy and eigenvector 
were performed using a Modified Lanczos algorithm\cite{gagliano1986correlation} (for details see SI).
The code was benchmarked by reproducing the results in \cite{chaloupka2013zigzag}. The chosen parameters were verified to be consistent with the zigzag ground state of Na$_{2}$IrO$_{3}$ 
by calculating 
structure factors $S(\overrightarrow{Q)}$ \cite{rau2014generic,rau2014trigonal,yadav2016kitaev}(see SI). 

\begin{table}
\begin{tabular}{|c|c|c|c|c|c|c|}
\hline 
Model & $J_{h}$ & $J_{K}$ & $J_{2}$ & $J_{3}$ & $\Gamma$ & $\Gamma^{\prime}$\tabularnewline
\hline 
\hline 
Antiferromagnetic Kitaev (\textit{Model A}) & - & + & $\times$ & $\times$ & $\times$ & $\times$\tabularnewline
\hline 
Ferromagnetic Kitaev (\textit{Model B}) & + & - & + & + & $\times$ & $\times$\tabularnewline
\hline 
Ferromagnetic Kitaev (\textit{Model C}) & + & - & $\times$ & $\times$ & + & -\tabularnewline
\hline 
\end{tabular}

\caption{Models considered for exact diagonalisation calculations, where $J_{h}$ refers to the nearest-neighbor Heisenberg interaction, $J_{K}$ refers to the Kitaev interaction, $J_{2}$ and $J_{3}$ refer to 
further-neighbor Heisenberg terms, and $\Gamma$ and $\Gamma^{\prime}$ refer to symmetric off-diagonal exchange interactions.}

\end{table}

\begin{figure}
\includegraphics[width=0.9\columnwidth]{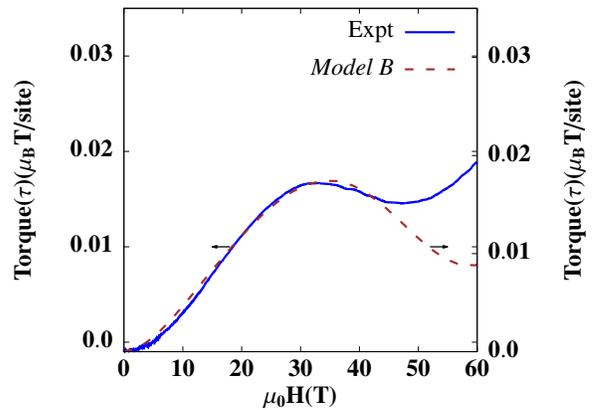}

\caption{\label{fig:poskh} Torque as a function of magnetic field (in $\mu_{B}$ tesla
per site) for \textit{Model B} (denoted by $\tau_{B}$) with parameters $J_{\rm{h}}=3.6$, $J_{\rm{K}}=-30.0$ (in meV),corresponding to the orientation
 $\theta=42^{\circ}$, $\phi=0^{\circ}$. In this case, further neighbor interactions $J_{2}=0.6$, $J_{3}=1.8$ (in meV) are necessary
to stabilize a zigzag ground state. The experimental data (solid line) for this orientation is plotted along with the torque
response (dashed line) calculated for this model for comparison.}
\end{figure}

\begin{figure}
\includegraphics[width=0.9\columnwidth]{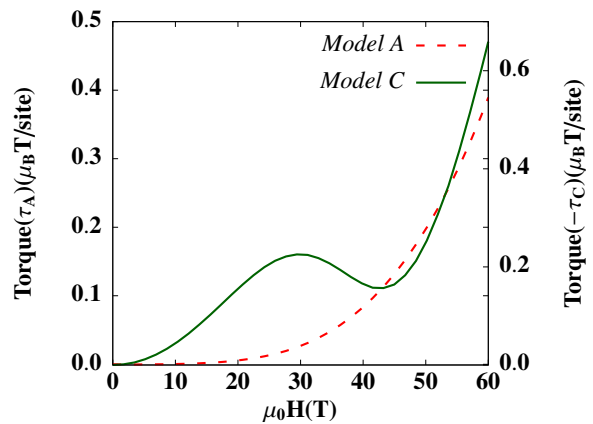}

\caption{\label{fig:negkh}Magnetic torque calculated as a function of magnetic field (in $\mu_{B}$ tesla
per site) for \textit{Model A} (denoted by $\tau_{A}$) with parameters $J_{\rm{h}}=-4.0$, $J_{\rm{K}}=21.0$ (in
meV),
corresponding to the orientation $\theta=36^{\circ}$, $\phi=0^{\circ}$,
and for \textit{Model C} (denoted by $\tau_{C}$) with parameters $J_{\rm{h}}=4.0$, $J_{\rm{K}}=-16.0,$ $\Gamma=2.4$,
$\Gamma^{\prime}=-3.2$ (in meV), corresponding to the same orientation. In \textit{Model A} (dashed line) characterized by a stable zigzag phase, 
no peak-dip feature appears, unlike experimental observations. In contrast, in \textit{Model C} (solid line), where a fine-tuned zigzag phase requires the introduction of nearest-neighbor
anisotropic terms $\Gamma$ and $\Gamma^{\prime}$, the magnetic field dependence of magnetic torque shows a peak-dip feature corresponding with experiment. }
\end{figure}

The calculated torque responses for the different models are shown in Figures \ref{fig:poskh}
and \ref{fig:negkh}. We find that the peak-dip feature in the torque response is reproduced only by \textit{Models B $\&$ C}), 
whereas \textit{Model A} displays a monotonic increase in the magnetic torque with magnetic field. We have performed exact diagonalization 
simulations for 
magnetic fields up to 300 T for \textit{Model A} (for the parameters used in Fig. \ref{fig:negkh}), and found a single peak in the torque response at a field slightly 
lower than 150 T, beyond which it
decreases with increase in field strength and no further features are observed. We have also considered variants of \textit{Model A} with isotropic $J_{2}$ and $J_{3}$ as well as anisotropic $\Gamma$ and 
$\Gamma^{\prime}$ terms, and have confirmed the absence of any peak-dip features even with such additional terms present (please refer to Table I in the SI for a 
summary of the different 
variants considered). 

Our results strongly indicate that Na$_2$IrO$_3$ is described by a model dominated by a ferromagnetic Kitaev exchange.  The distinctive peak-dip feature in the torque response provides an 
independent handle for constraining experimental data. We note that classical Monte Carlo simulations were unable to reproduce the feature, underlining the importance of quantum effects 
in this material, as has also been emphasized in the recent literature \cite{janssen2017magnetization}. Of the two types of ferromagnetic Kitaev exchange models we consider, in \textit{Model B}, 
the peak-dip feature is observed over a large parameter range, while in \textit{Model C}, it only appears upon inclusion of a significant $\Gamma^{\prime} < 0$ term, which is physically  
associated with trigonal distortion in Na$_{2}$IrO$_{3}$. The inclusion of significant anisotropy terms in \textit{Model B} does not yield additional peak-dip features,
and the feature survives only for relatively small values of additional anisotropic interactions. \textit{Models B} and \textit{C} can thus potentially be distinguished by high 
field torque magnetometry measurements 
on chemically doped Na$_2$IrO$_3$ with various extents of trigonal distortion.

\begin{figure}
\includegraphics[width=1.0\columnwidth,height=0.8\columnwidth]{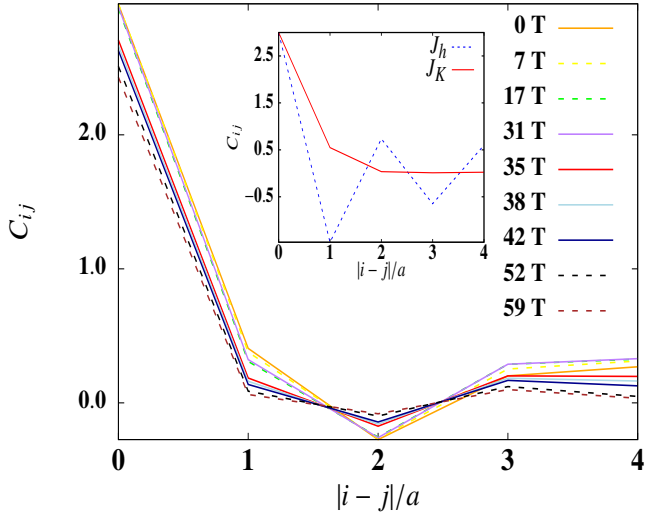}

\caption{\label{fig:cij}The correlation functions $C_{ij}$ calculated as
a function of $\frac{|i-j|}{a}$, $a$ being the distance between two
neighboring sites, with parameters $J_{\rm{h}}=4.0$, $J_{\rm{K}}=-16.0$, $\Gamma=2.4$,
$\Gamma^{\prime}=-3.2$ (in meV), for an orientation of $\theta=36^{\circ}$, $\phi=0^{\circ}$.
The inset shows the corresponding plots for a pure Heisenberg model
with $J_{\rm{h}}=16.0$ meV(blue) and a pure Kitaev model with $J_{\rm{K}}=-16.0$
meV(red). It can be clearly seen that for higher fields(>30 T), the
correlation functions fall rapidly with distance, signalling spin-liquid physics. }
\end{figure}

We compute the evolution of the spin correlation functions with 
distance for increasing magnetic field values. The extent of decay of the correlation functions with distance reveals the presence or absence of long range correlations in the high field regime. The correlation
functions $C_{ij}=<(\overrightarrow{\sigma_{i}}-<\overrightarrow{\sigma_{i}}>).(\overrightarrow{\sigma_{j}}-<\overrightarrow{\sigma_{j}}>)>$ are calculated for a chosen set of neighboring sites in the 
24-site cluster, and
plotted in Fig.~\ref{fig:cij} as a function of $\frac{|i-j|}{a}$ ($a$ being the distance between nearest neighbor sites) for different values of the applied magnetic field. We find that the decay of the 
correlation functions is much faster at relatively higher values of the applied field, and the amplitude of their oscillation falls off rapidly with increasing fields,
 in particular above the zigzag ordering scale. Furthermore, structure factor calculations do not show a crossover from antiferromagnetic zigzag order to any of the known ordered states at the position of the
 metamagnetic transition manifested through the peak-dip in the transverse magnetization. Indications are therefore that the high field regime beyond the peak-dip feature manifests spin-liquid physics
 in Na$_{2}$IrO$_{3}$. 
 
Our work sheds light on the universality of field-induced spin liquid physics in Kitaev systems, which we find to be signalled by
a peak dip structure in the anisotropic 
magnetisation at the zigzag ordering scale both in Na$_{2}$IrO$_{3}$ and $\alpha-$RuCl$_{3}$ \cite{sears2017phase,yadav2016kitaev,PhysRevLett.119.227208,PhysRevB.95.241112,PhysRevB.92.235119}.
Recent calculations of thermal Hall effect in  $\alpha-$RuCl$_{3}$~\cite{Cookmeyer2018spin} as well as 
electron energy loss spectroscopy experiments~\cite{koitzsch2017low}
also favour a dominant ferromagnetic Kitaev model. The striking similarities between these two materials indicates that the experimental features in the magnetoresponse 
that we report here are governed by intrinsic Kitaev physics, and not peculiarities associated with parameters beyond the scope of our model such as interlayer couplings and disorder 
characteristics, expected to be very different for these two materials. The microscopic models we calculate here are thus indicated to be relevant to a broad class of honeycomb 
Kitaev materials for exploring a field-induced spin liquid phase.

\begin{figure}
\includegraphics[width=0.8\columnwidth]{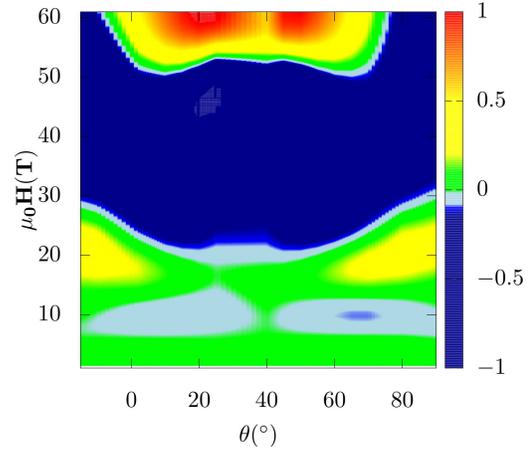}

\caption{\label{fig:cheshire} {The figure shows a calculated contourplot of $\frac{d\tau}{dH}$
in the $\theta-H$ plane, for parameters $J_{\rm{h}}=3.6$, $J_{\rm{K}}=-18.0$, $J_{2}=2.4$ and $J_{3}=1.8$ (in meV),
i.e. \textit{Model B},
corresponding to the azimuthal angle $\phi=20^{\circ}$.
We find that the position of the peak-dip
feature, indicated by the regions where $\frac{d\tau}{dH}$ changes
sign, shifts closer to the origin for increasing(decreasing) values
of the polar angle $\theta$ for $\theta$ close to $0^{\circ}$ ($90^{\circ}$),
in agreement with the experimental results. At the extreme values of $\theta$, the width of the region of
nonmonotonicity increases, at variance
with experiment. The torque values obtained in our simulations can be negative, and in
such cases we plot $-\frac{d\tau}{dH}$ instead. }
}
\end{figure}


The authors gratefully acknowledge useful discussions with Giniyat Khaliullin, Itamar Kimchi, Subhro Bhattacharjee and Steve Winter. We thank E. V. Sampathkumaran for the generous 
use of facilities for crystal growth. 
VT acknowledges DST for a Swarnajayanti grant (No. DST/SJF/PSA-0212012-13). SDD and SES acknowledge support from the Royal Society, the Winton Programme for the Physics of Sustainability, 
and the European Research 
Council under the European Unions
Seventh Framework Programme (grant number FP/2007-2013)/ERC Grant Agreement number 337425. LB is supported by DOE-BES through award DE-SC0002613. GC acknowledges the support of the US National Science 
Foundation via 
grant DMR 1712101. AM acknowledges the support of the HFML-RU/FOM, member of the European Magnetic Field Laboratory(EMFL). HYK acknowledges 
the support of the NSERC
of Canada and the center for Quantum Materials at the University of
Toronto. 

High field experiments were performed by S.D.D. with contributions from Z.Z., E.M., R.D.M., G.L., L.B. and A.M.. Theory was developed and calculations performed by S.K. and V.T. with contributions 
from H.Y.K. and J.G.R. Single crystal growth was performed by S.D.D. and G.C. The project was conceived and supervised by S.E.S. and V.T. The manuscript was written by S.D.D., S.K., V.T., and S.E.S. 
with inputs from all the authors.

\bibliographystyle{apsrev4-1}


\author{Sitikantha D. Das$^{1,11}$}
\author{Sarbajaya Kundu$^{2}$}
\author{Zengwei Zhu$^{3}$}
\altaffiliation[Now at: ]{National High Magnetic Field Center and School of Physics, Huazhong University of Science and Technology, Wuhan 430074, China}
\author{Eundeok Mun$^{3}$}
\altaffiliation[Now at: ]{Department of Physics, Simon Fraser University, Burnaby, BC, Canada V5A-1S6}
\author{Ross D. McDonald$^{3}$}
\author{Gang Li$^{4}$}
\altaffiliation[Now at: ]{Department of Physics, University of Michigan, 500 S State St, Ann Arbor, MI 48109, USA}
\author{Luis Balicas$^{4}$}
\altaffiliation[Now at: ]{Institute of Physics, Chinese Academy of Sciences, P.O Box 603, Beijing 100190, China}
\author{Alix McCollam$^{5}$}
\author{Gang Cao$^{6,7}$}
\author{Jeffrey G. Rau$^{8}$}
\author{Hae-Young Kee$^{9,10}$}
\author{Vikram Tripathi$^{2}$}
\author{Suchitra E. Sebastian$^{1}$}

\affiliation{$^{1}$Cavendish Laboratory, University of Cambridge, J J Thomson Avenue, Cambridge CB3 0HE, UK}
\affiliation{$^{2}$Department of Theoretical Physics, Tata Institute of Fundamental Research, Homi Bhabha Road, Colaba, Mumbai 400005, India}
\affiliation{$^{3}$Los Alamos National Laboratory, Los Alamos, New Mexico 87545, USA}
\affiliation{$^{4}$National High Magnetic Field Laboratory,1800 E. Paul Dirac Drive, Tallahassee, FL 32310, USA}
\affiliation{$^{5}$High Field Magnet Laboratory (HFML - EMFL), Radboud University, 6525 ED, Nijmegen, The Netherlands}
\affiliation{$^{6}$Center for Advanced Materials and Department of Physics and Astronomy, University of Kentucky, Lexington, Kentucky 40506, USA}
\affiliation{$^{7}$Department of Physics, 390 UCB, University of Colorado, Boulder, CO 80309, USA}
\affiliation{$^{8}$Department of Physics and Astronomy, University of Waterloo, Ontario, N2L 3G1, Canada}
\affiliation{$^{9}$Department of Physics, University of Toronto, Toronto, Ontario M5S 1A7, Canada}
\affiliation{$^{10}$Canadian Institute for Advanced Research/Quantum Materials Program, Toronto, Ontario MSG 1Z8, Canada}
\affiliation{$^{11}$Department of Physics, IIT, Kharagpur, Kharagpur 721302, India}


\section*{A1. Experimental details}

Crystals of Na$_{2}$IrO$_{3}$ about 100\textbf{ }micron along a
side were prepared using Na$_{2}$CO$_{3}$ slightly in excess \cite{singh2010antiferromagnetic}.
The insulating nature of the sample was confirmed using four-probe
resistivity measurements. Magnetization measurements were performed
using a Quantum Design SQUID magnetometer upto a field of 5 tesla,
and these showed the presence of an antiferromagnetic transition around
$T_{N}\sim$15 K, below which the system orders into a zigzag phase
\cite{choi2012spin,ye2012direct,liu2011long}. A linear Curie-Weiss
fit to the high temperature inverse susceptibility data gives an effective
Ir moment $\mu_{eff}=1.67\mu_{B}$ and Curie temperature $\theta_{p}=-116$
K. This gives a frustration index $\frac{\theta_{p}}{T_{N}}\sim8$,
which is in agreement with the results obtained from other groups.

The torque ($\overrightarrow{\tau}=\overrightarrow{m}\times\overrightarrow{B}$)
was measured using a PRC 120 piezoresistive cantilever at the pulsed
field facility at the National High Magnetic Field Laboratory, Los
Alamos. The crystal was mounted on the cantilever with vacuum grease.\textbf{
}The cantilever assembly was mounted on a sample holder made of G-10
and capable of rotation about the magnetic field axis. The pulse duration
was 25 ms. All the angles ($\theta$) were measured with respect to
the normal to the flat surface of the crystal (which is the nominal
c-axis) and the magnetic field $H$. For each value of $\theta$ the
torque response was measured for the increasing and decreasing cycles
of the pulse which had a high degree of overlap ruling out significant
magneto-caloric effects which might become evident in pulsed field
measurements. Further measurements were performed for various in-planar
orientations $\phi$ of the crystal. The torque response for a second
crystal from an independently grown batch, mentioned in the main text,
is shown in Fig. \ref{fig:secondcrystal}.

\begin{figure}
\includegraphics[width=0.8\columnwidth]{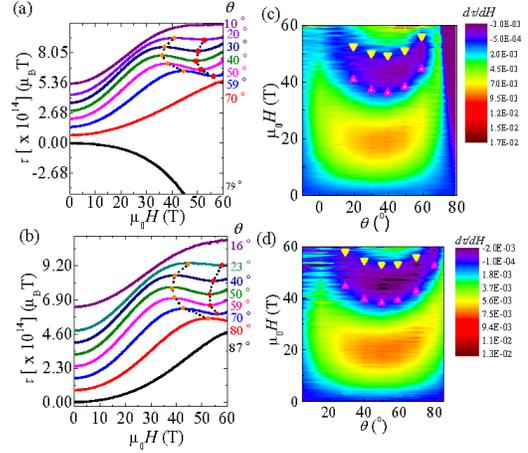}

\caption{\label{fig:secondcrystal}Torque as a function of magnetic field($H$)
measured independently for a second crystal corresponding to two different
in-plane orientations $\phi$ separated by $90^{\circ}$, is shown
in (a) and (b), and is again found to show nonmonotonous behavior
for a range of orientations. The corresponding plots for $\frac{d\tau}{dH}$
as a function of $H$ are shown in (c) and (d). $\theta$ is the angle
that the normal to the crystal makes with the magnetic field, and
is defined in the main text. }
\end{figure}

\begin{figure}
\includegraphics[width=0.5\columnwidth]{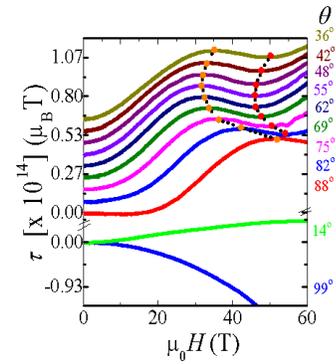}

\caption{\label{fig:secondcrystal}Torque ($\tau$) as a function of magnetic
field for different angular orientations ($\theta$) and $\phi=0^{\circ}$
corresponding to the first crystal as mentioned in the main text. }
\end{figure}

\section*{A2. Numerical setup and Exact Diagonalization algorithm}

\begin{figure}
\includegraphics[width=0.6\columnwidth]{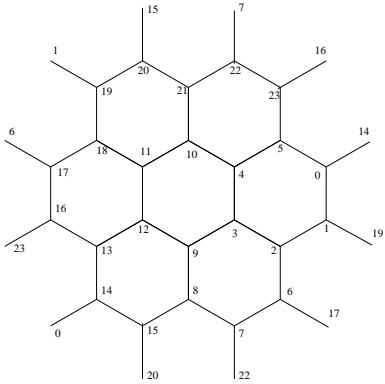}

\caption{\label{fig:kfrag}Our 24-site fragment with periodic boundary conditions}
\end{figure}

The N lattice sites were numbered 0,1,2...N-1 (for N=24), and specific
pairs of these sites were identified as `bonds' or `links', of type
$x$, $y$ or $z$. Every site has a spin with two possible states
$|1>$ or $|0>$. The system then has $2^{N}$ configurations or underlying
basis states, where each configuration is denoted by $|s_{N-1},s_{N-2}...s_{0}>$
with $s_{i}=0,1$. Corresponding to such a set of binary numbers,
we have a decimal equivalent given by $|s_{N-1}{\rm {x}2^{N-1}+s_{N-2}{\rm {x}2^{N-2}+...s_{0}{\rm {x}2^{0}>}}}$.
The basis vectors were thus denoted as $|0>$, $|1>$...$|2^{N-1}>$.
An arbitrary state vector $|\psi>$ can be expanded in terms of these
basis vectors as $|\psi>=\sum_{i=0}^{2^{N-1}}a_{i}|i>$. The ground
state $\Psi_{0}$ for this Hamiltonian $H_{0}$ was determined using
the Modified Lanczos algorithm.

The Modified Lanczos algorithm \cite{gagliano1986correlation} requires
the initial selection of a trial vector $\psi_{0}$ (constructed using
a random number generator in our case) which should have a nonzero
projection on the true ground state of the system in order for the
algorithm to converge properly. A normalized state $\psi_{1}$, orthogonal
to $\psi_{0}$, is defined as

\begin{equation}
\psi_{1}=\frac{H_{0}\psi_{0}-<H_{0}>\psi_{0}}{\sqrt{<H_{0}^{2}>-<H_{0}>^{2}}}\label{eq:2}
\end{equation}

In the basis $\{\psi_{0},\psi_{1}\}$, $H_{0}$ has a 2x2 representation
which is easily diagonalized. Its lowest eigenvalue and corresponding
eigenvector are better approximations to the true ground state energy
and wavefunction than the quantities $<H_{0}>$ and $\psi_{0}$ considered
initially. The improved energy and wavefunction are given by

\begin{equation}
\epsilon=<H_{0}>+b\alpha\label{eq:3}
\end{equation}

and 
\begin{equation}
\tilde{\psi_{0}}=\frac{\psi_{0}+\alpha\psi_{1}}{\sqrt{1+\alpha^{2}}}\label{eq:4}
\end{equation}

where $b=\sqrt{<H_{0}^{2}>-<H_{0}>^{2}}$, $f=\frac{<H_{0}^{3}>-3<H_{0}><H_{0}^{2}>+2<H_{0}>^{3}}{2b^{3}}$
and $\alpha=f-\sqrt{1+f^{2}}$. The method can be iterated by considering
$\tilde{\psi_{0}}$ as a new trial vector and repeating the above
steps. The Modified Lanczos method helps in obtaining a reasonably
good approximation to the actual ground state of the system while
storing only three vectors, $\psi_{0}$, $H_{0}\psi_{0}$ and $H_{0}^{2}\psi_{0}$
rather than the entire Hamiltonian in the spin basis representation.
This is especially advantageous as the number of basis vectors increases
rapidly with the number of spin sites. In the regular Lanczos algorithm,
the matrix is first reduced to a tridiagonal form before computing
the ground state eigenvector. However, there can be issues with the
convergence to the true ground state because of loss of orthogonality
among the vectors. This is circumvented in this algorithm as orthogonality
is enforced at each and every step of the iteration.

After the determination of the ground state $\Psi_{0}$ to a reasonable
approximation, the magnetization $\vec{m}_{|\psi>}$ was obtained
in this state with components ($m_{x},m_{y},m_{z}$), where $m_{\gamma}=<\psi|\sum_{i=0}^{N-1}\sigma_{i}^{\gamma}|\psi>$,
and was transformed to the lab frame from the octahedral frame, the
components in the lab frame being ($m_{X},m_{Y},m_{Z}$). Finally,
in the lab frame, torque $\Gamma_{X}=m_{Y}H$.

\section*{A3. Coordinate system transformations}

For our exact diagonalization calculations, we have transformed the
external magnetic field from the laboratory frame to the $\rm{IrO}_{6}$ octahedral
frame by defining intermediate crystal and cantilever axes, and transformed
the calculated magnetization back from this frame to the lab frame.
We explain the transformations used in the following:

\subsection*{Notations:}

\emph{Laboratory axes}: $\hat{X},\hat{Y},\hat{Z}$

\paragraph*{\textmd{Cantilever axes: $\hat{x,}\hat{y},\hat{z}$}}

\paragraph*{\textmd{Crystal axes: $\hat{a},\hat{b},\hat{c}$}}

\paragraph*{\textmd{Octahedral axes: $\hat{p},\hat{q},\hat{r}$}}

\subsection*{Laboratory to cantilever axes:}

The lab $\hat{X}$-axis and the cantilever $\hat{x}$-axis are always
coincident. Let $\theta$ be the angle between the $\hat{Z}$ and
$\hat{z}$ axes.We have
\[
|\hat{x}\hat{y}\hat{z}>=M_{Lab\rightarrow Canti}|\hat{X}\hat{Y}\hat{Z}>
\]
\[
|\hat{X}\hat{Y}\hat{Z}>=L_{Canti\rightarrow Lab}|\hat{x}\hat{y}\hat{z}>
\]
where
\[
M_{Lab\rightarrow Canti}=\left(\begin{array}{ccc}
1 & 0 & 0\\
0 & \cos\theta & -\sin\theta\\
0 & \sin\theta & \cos\theta
\end{array}\right)
\]
\[
L_{Canti\rightarrow Lab}=\left(\begin{array}{ccc}
1 & 0 & 0\\
0 & \cos\theta & \sin\theta\\
0 & -\sin\theta & \cos\theta
\end{array}\right)
\]

\subsection*{Cantilever to crystal axes:}

The honeycomb layer formed by the Ir atoms resides on the crystallographic
$ab$ plane. Let the $\hat{a}$-axis of the crystal make an angle $\phi$
with the $\hat{x}$-axis of the cantilever. Then,
\[
|\hat{a}\hat{b}\hat{c}>=M_{Canti\rightarrow Crystal}|\hat{x}\hat{y}\hat{z}>
\]
\[
|\hat{x}\hat{y}\hat{z}>=L_{Crystal\rightarrow Canti}|\hat{a}\hat{b}\hat{c}>
\]
\[
M_{Canti\rightarrow Crystal}=\left(\begin{array}{ccc}
\cos\phi & \sin\phi & 0\\
-\sin\phi & \cos\phi & 0\\
0 & 0 & 1
\end{array}\right)
\]
\[
L_{Crystal\rightarrow Canti}=\left(\begin{array}{ccc}
\cos\phi & -\sin\phi & 0\\
\sin\phi & \cos\phi & 0\\
0 & 0 & 1
\end{array}\right)
\]

\subsection*{Crystal to octahedral axes:}

Since the {[}111{]} direction in the octahedral frame is perpendicular
to the honeycomb lattice, the unit vectors are related as follows:
\[
\hat{c}=\frac{\hat{p}+\hat{q}+\hat{r}}{\sqrt{3}}
\]
\[
\hat{b}=\frac{-\hat{p}+\hat{q}}{\sqrt{2}}
\]
\[
\hat{a}=\frac{\hat{p}+\hat{q}-2\hat{r}}{\sqrt{6}}
\]

Then,
\[
|\hat{p}\hat{q}\hat{r}>=M_{Crystal\rightarrow Octa}|\hat{a}\hat{b}\hat{c}>
\]
\[
|\hat{a}\hat{b}\hat{c}>=L_{Octa\rightarrow Crystal}|\hat{p}\hat{q}\hat{r}>
\]
\[
M_{Crystal\rightarrow Octa}=\left(\begin{array}{ccc}
\frac{1}{\sqrt{6}} & -\frac{1}{\sqrt{2}} & \frac{1}{\sqrt{3}}\\
\frac{1}{\sqrt{6}} & \frac{1}{\sqrt{2}} & \frac{1}{\sqrt{3}}\\
-\sqrt{\frac{2}{3}} & 0 & \frac{1}{\sqrt{3}}
\end{array}\right)
\]
\[
L_{Octa\rightarrow Crystal}=\left(\begin{array}{ccc}
\frac{1}{\sqrt{6}} & \frac{1}{\sqrt{6}} & -\sqrt{\frac{2}{3}}\\
-\frac{1}{\sqrt{2}} & \frac{1}{\sqrt{2}} & 0\\
\frac{1}{\sqrt{3}} & \frac{1}{\sqrt{3}} & \frac{1}{\sqrt{3}}
\end{array}\right)
\]

\begin{widetext}
\subsection*{Lab to octahedral and octahedral to lab frame:}

Let the components of the magnetic field be $(0,0,H)$ in the lab
frame and $(h_{p},h_{q},h_{r})$ in the octahedral frame. Then
\[
|h_{p}h_{q}h_{r}>=M_{Crystal\rightarrow Octa}M_{Canti\rightarrow Crystal}M_{Lab\rightarrow Canti}|00H>
\]
which finally gives us
\[
h_{p}=(-\frac{1}{\sqrt{6}}\sin\theta\sin\phi+\frac{1}{\sqrt{2}}\sin\theta\cos\phi+\frac{1}{\sqrt{3}}\cos\theta)H
\]
\[
h_{q}=(-\frac{1}{\sqrt{6}}\sin\theta\sin\phi-\frac{1}{\sqrt{2}}\sin\theta\cos\phi+\frac{1}{\sqrt{3}}\cos\theta)H
\]
\[
h_{r}=(\sqrt{\frac{2}{3}}\sin\theta\sin\phi+\frac{1}{\sqrt{3}}\cos\theta)H
\]
Let the components of the magnetization vector $\overrightarrow{m}$
be $(m_{X},m_{Y},m_{Z})$ in the lab frame and $(m_{p},m_{q},m_{r})$
in the octahedral frame. Then,
\[
|m_{X}m_{Y}m_{Z}>=L_{Canti\rightarrow Lab}L_{Crystal\rightarrow Canti}L_{Octa\rightarrow Crystal}|m_{p}m_{q}m_{r}>
\]
from where we find 
\[
m_{Z}=(-\frac{m_{p}}{\sqrt{6}}-\frac{m_{q}}{\sqrt{6}}+m_{r}\sqrt{\frac{2}{3}})\sin\theta\sin\phi+\frac{m_{p}-m_{q}}{\sqrt{2}}\sin\theta\cos\phi+\frac{(m_{p}+m_{q}+m_{r})}{\sqrt{3}}\cos\theta
\]
and
\[
m_{Y}=(\frac{m_{p}}{\sqrt{6}}+\frac{m_{q}}{\sqrt{6}}-m_{r}\sqrt{\frac{2}{3}})\cos\theta\sin\phi+\frac{(-m_{p}+m_{q})}{\sqrt{2}}\cos\theta\cos\phi+\frac{(m_{p}+m_{q}+m_{r})}{\sqrt{3}}\sin\theta
\]
\end{widetext}

\begin{figure*}
(a)\includegraphics[width=0.55\columnwidth]{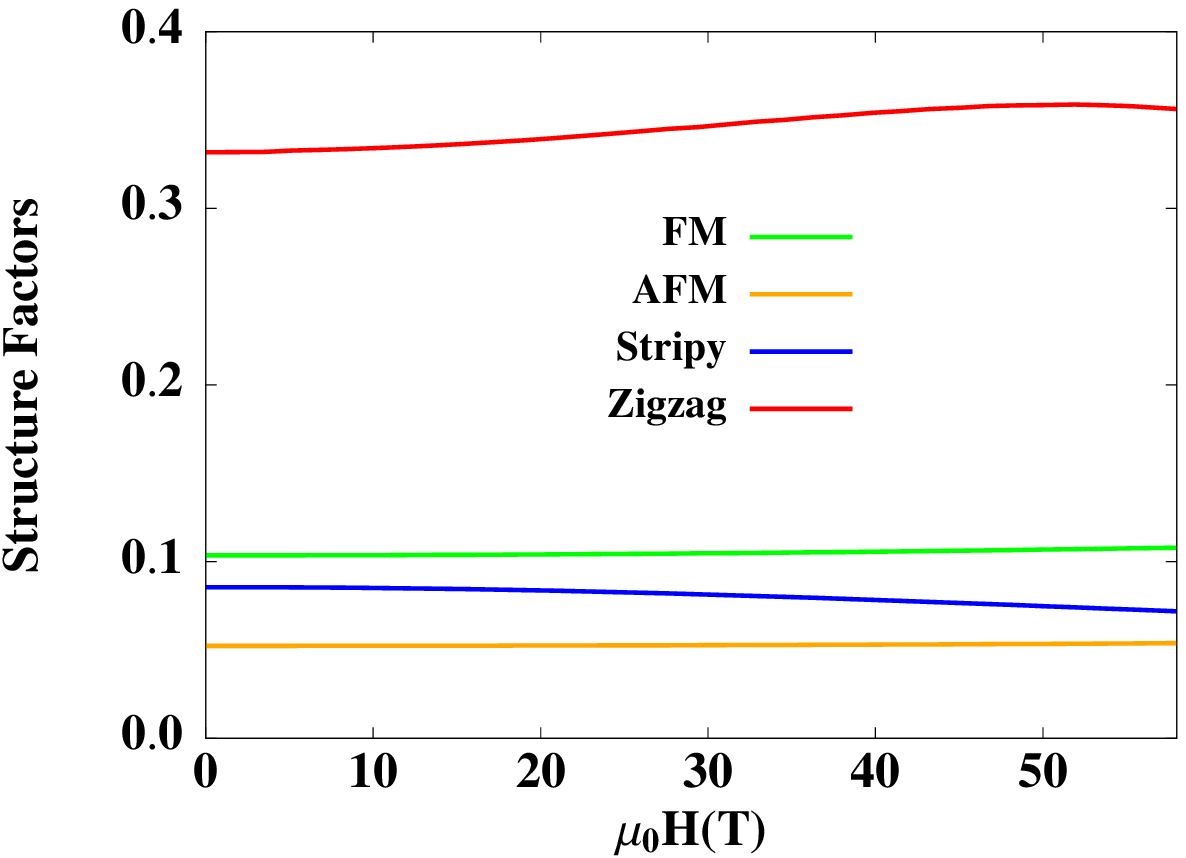}(b)\includegraphics[width=0.55\columnwidth]{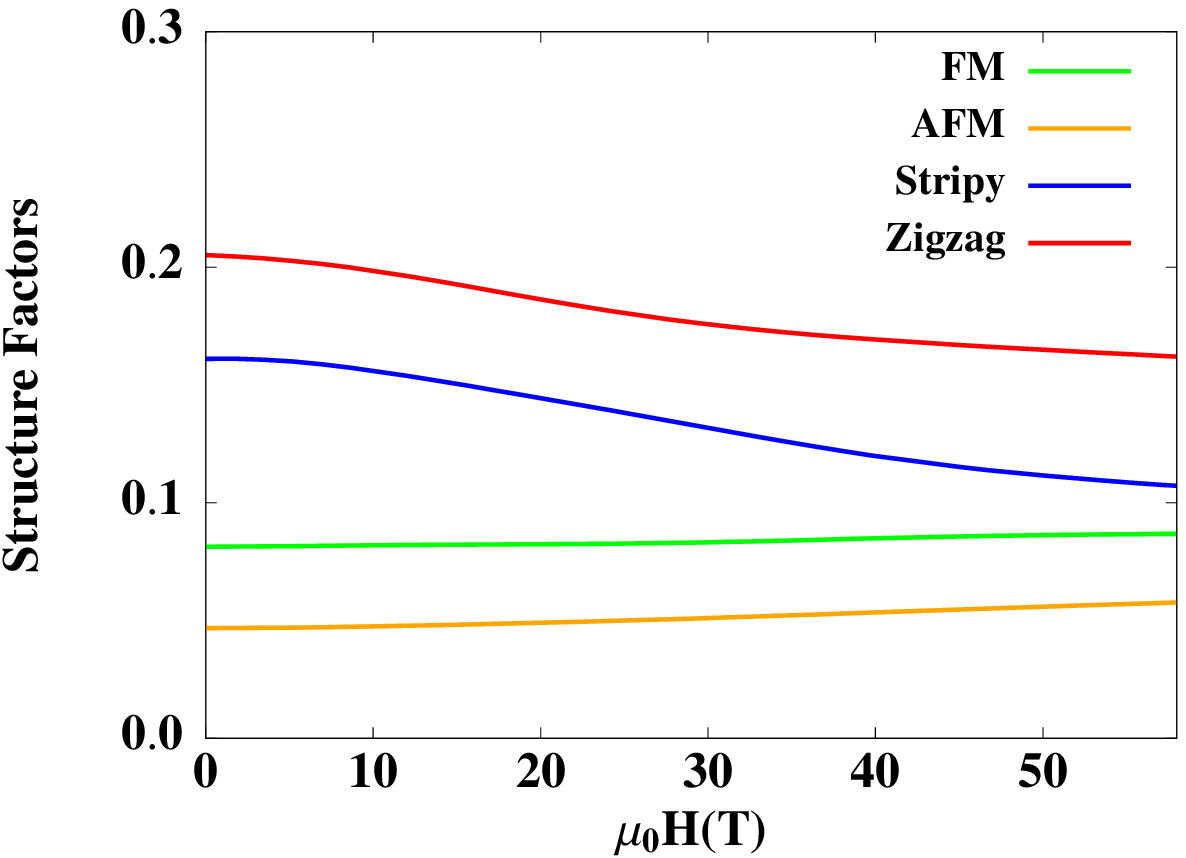}
(c)\includegraphics[width=0.55\columnwidth]{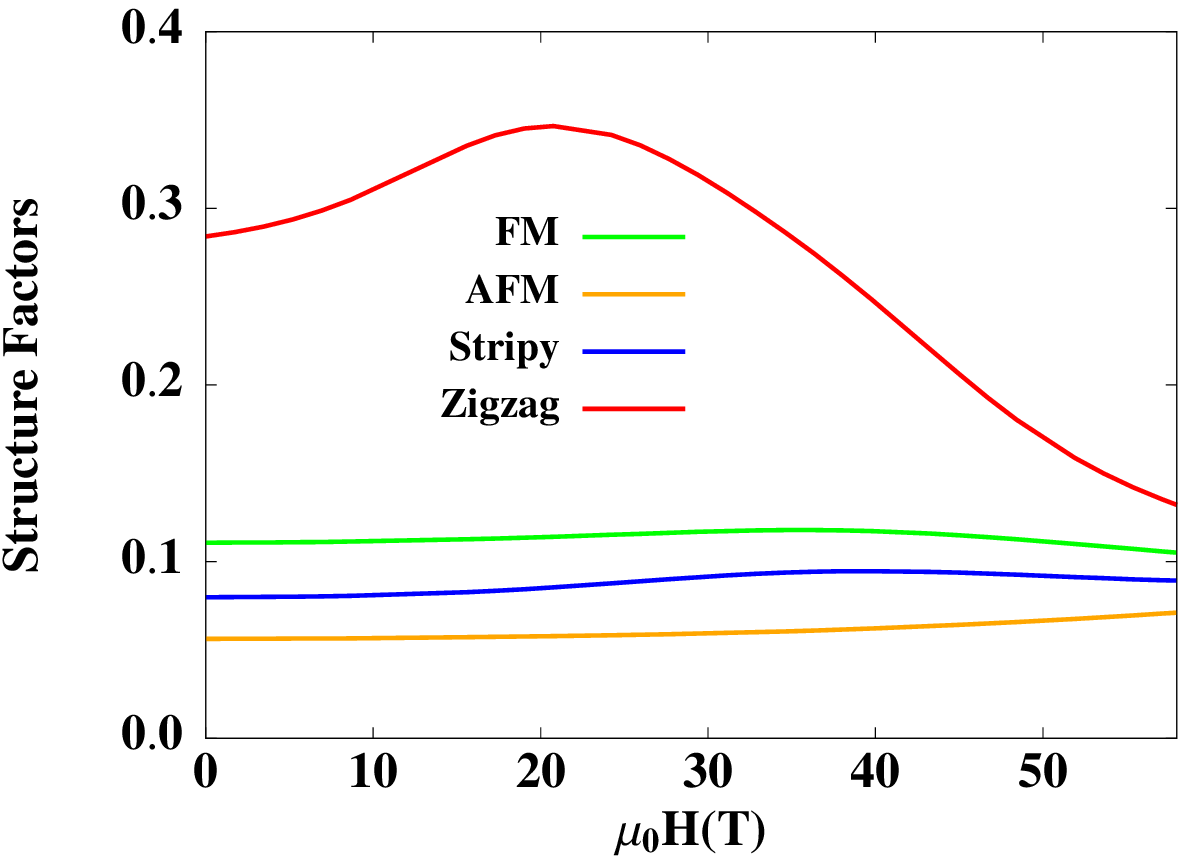}

\caption{\label{fig:sf}Evolution of structure factors for different ordered
phases as a function of the field for (a) $J_{h}=3.2$, $J_{\rm{K}}=-12$,
$J_{2}=4$, $J_{3}=2$(in meV), (b) $J_{h}=1.6$, $J_{\rm{K}}=-16.0$,
$J_{2}=1.2$ and $J_{3}=0.8$(in meV), and (c) $J_{h}=4.0$, $J_{\rm{K}}=-16.0$,
$\Gamma=2.4$ and $\Gamma^{\prime}=-3.2$(in meV). }
 
\end{figure*}

\section*{A4. Structure factor calculations }

To determine different phases of the system in the presence of an
applied magnetic field, one needs to calculate the adapted structure
factors acting as order parameters \cite{trousselet2011effects}. The
corresponding dominant order wavevectors $\overrightarrow{Q}=\overrightarrow{Q_{max}}$
characterize the nature of the magnetic ordering in various field
regimes. The static structure factors $S(\overrightarrow{Q})$ for
different spin configurations are given by
\begin{widetext}
\begin{equation}
S_{zigzag}^{\gamma}=\frac{1}{N^{2}}\sum_{r,r^{\prime},\beta,\beta^{\prime}}\exp[i\overrightarrow{Q_{\gamma}.}(\overrightarrow{r^{\prime}}-\overrightarrow{r})]\nu_{\beta,\beta^{\prime}}
(<\overrightarrow{\sigma_{r,\beta}}.\overrightarrow{\sigma_{r^{\prime},\beta^{\prime}}}>-\sum_{\gamma}<\sigma_{r,\beta}^{\gamma}><\sigma_{r^{\prime},\beta^{\prime}}^{\gamma}>)\label{eq:3-1}
\end{equation}
\begin{equation}
S_{Neel}=\frac{1}{N^{2}}\sum_{r,r^{\prime},\beta,\beta^{\prime}}\nu_{\beta,\beta^{\prime}}
(<\overrightarrow{\sigma_{r,\beta}}.\overrightarrow{\sigma_{r^{\prime},\beta^{\prime}}}>-\sum_{\gamma}<\sigma_{r,\beta}^{\gamma}><\sigma_{r^{\prime},\beta^{\prime}}^{\gamma}>)\label{eq:4-1}
\end{equation}
\begin{equation}
S_{FM}=\frac{1}{N^{2}}\sum_{r,r^{\prime},\beta,\beta^{\prime}}
(<\overrightarrow{\sigma_{r,\beta}}.\overrightarrow{\sigma_{r^{\prime},\beta^{\prime}}}>-\sum_{\gamma}<\sigma_{r,\beta}^{\gamma}><\sigma_{r^{\prime},\beta^{\prime}}^{\gamma}>)\label{eq:5}
\end{equation}
\begin{equation}
S_{stripy}^{\gamma}=\frac{1}{N^{2}}\sum_{r,r^{\prime},\beta,\beta^{\prime}}\exp[i\overrightarrow{Q_{\gamma}.}(\overrightarrow{r^{\prime}}-\overrightarrow{r})]
(<\overrightarrow{\sigma_{r,\beta}}.\overrightarrow{\sigma_{r^{\prime},\beta^{\prime}}}>-\sum_{\gamma}<\sigma_{r,\beta}^{\gamma}><\sigma_{r^{\prime},\beta^{\prime}}^{\gamma}>)\label{eq:6}
\end{equation}
\end{widetext}
where each site is labeled by an index $i$ and a position in the unit
cell $\overrightarrow{r}$ ,$\beta$ denotes the sublattice index($\beta=A,B$), and $\gamma=x,y$ or $z$.
The contribution to the structure factors coming from the alignment
of the spins with the field direction has explicitly been deducted
in this definition. The structure factors for
the four different phases are plotted as a function of field for different
models in Fig. \ref{fig:sf}, which clearly shows that AFM zigzag
is the dominant spin configuration in all cases.

\section*{A5. Magnetization as a function of field}

\begin{figure}
(a)\includegraphics[width=0.45\columnwidth]{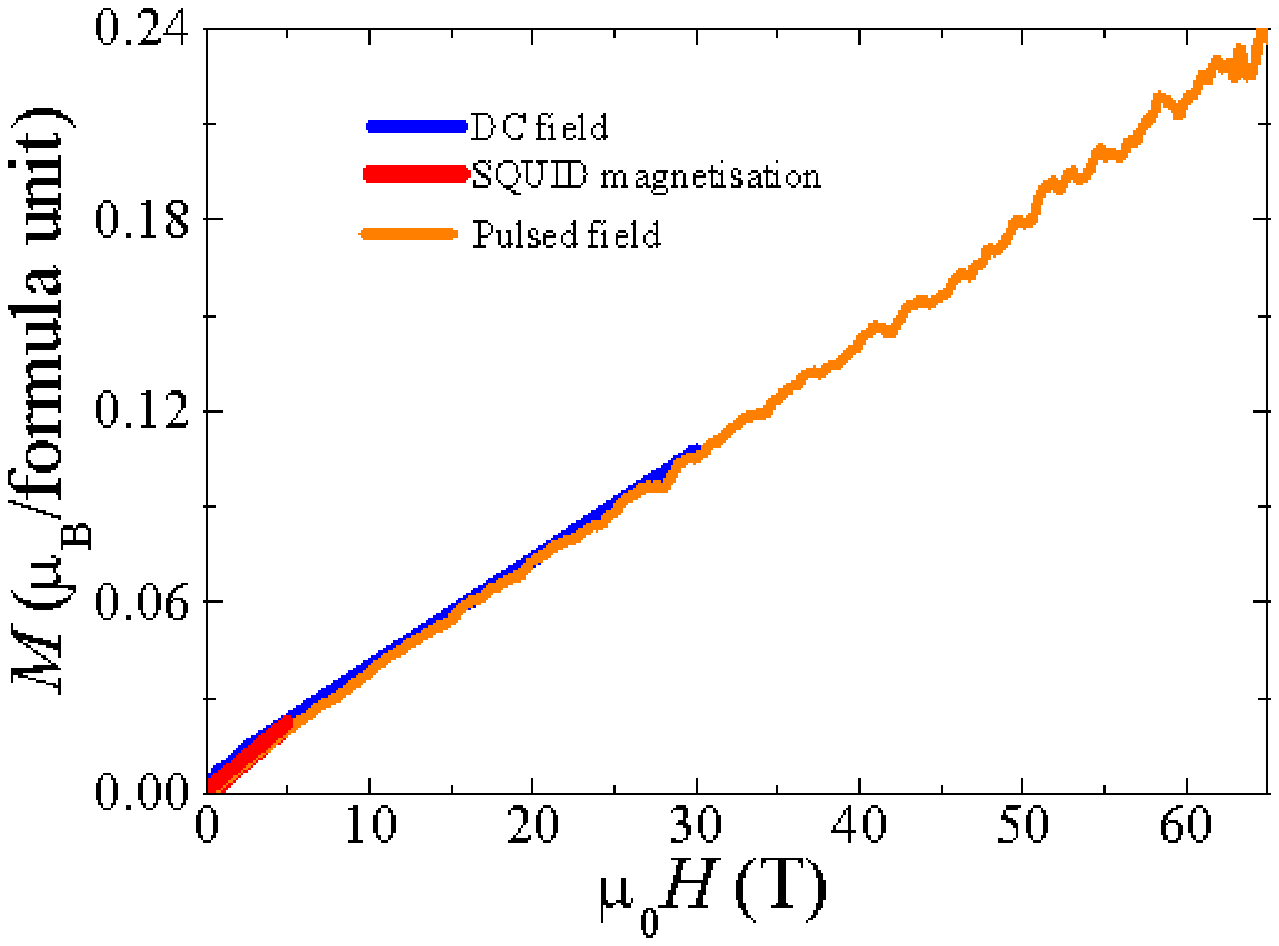}(b)\includegraphics[width=0.45\columnwidth]{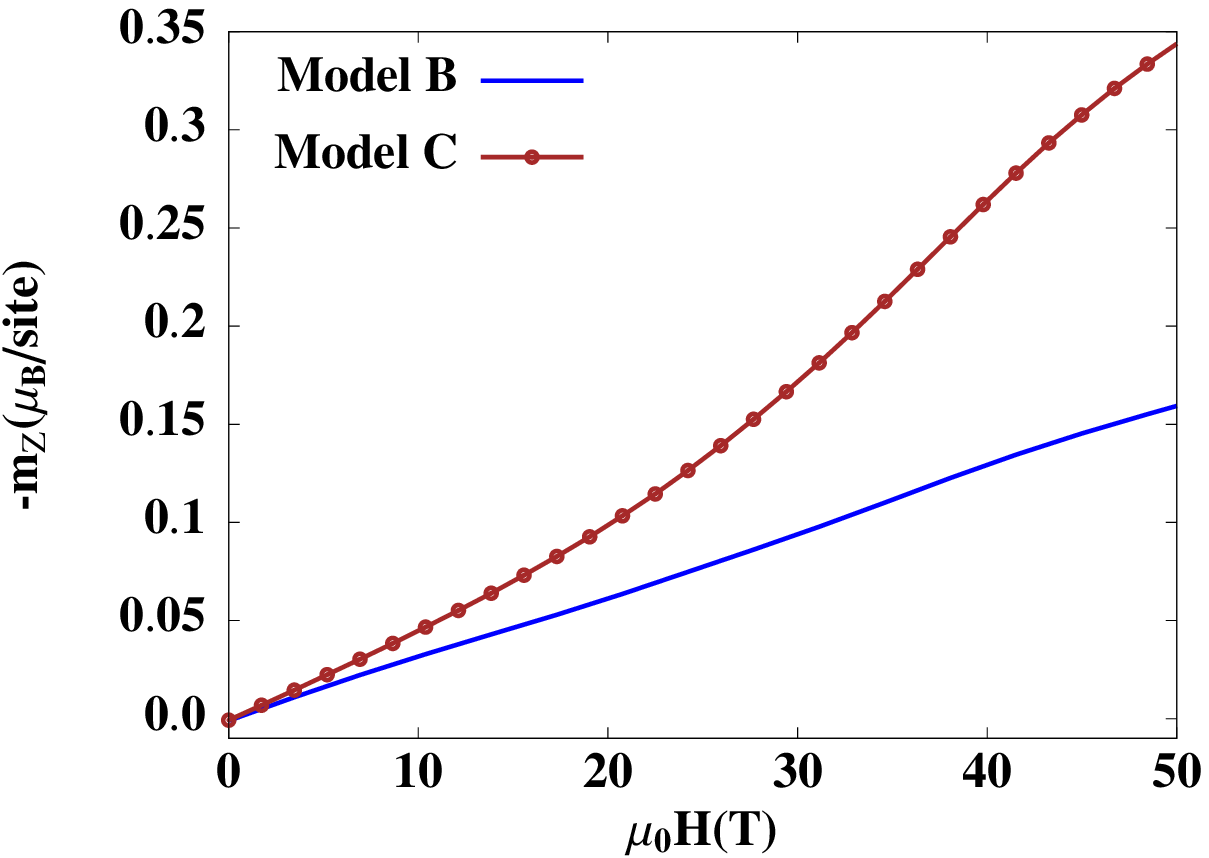}

\caption{\label{fig:mz}(a) Isotropic magnetization measured using an extraction
magnetometer in pulsed magnetic fields, and using a force magnetometer
in DC fields, shows no features up to 60 T, calibration is performed
using magnetization measurements on a pellet of sodium iridate in
a SQUID magnetometer, (b) Isotropic magnetization $m_{Z}$ (in $\mu_{B}$
per atom) calculated as a function of field, for model B with $J_{h}=2.4$,
$J_{{\rm {K}}}=-12.0$, $J_{2}=1.6$, $J_{3}=1.2$(in meV) for the
orientation $\theta=18^{\circ}$, $\phi=90^{\circ}$, and for model
C with $J_{h}=4.0$, $J_{{\rm {K}}}=-16.0$, $\Gamma=2.4$ and $\Gamma^{\prime}=-3.2$(in
meV), for the orientation $\theta=36^{\circ}$, $\phi=0^{\circ}$. }
\end{figure}

The isotropic magnetization($m_{Z}$) was measured using an extraction
magnetometer in pulsed magnetic fields up to 60 T and calibrated to
obtain $m_{Z}$ per site using force magnetometry measurements in
steady magnetic fields, and magnetization measurements in a SQUID
magnetometer. It is found to be largely featureless and increases
linearly with field up to 60 T. We have determined the behavior of
$m_{Z}$ per site numerically for different relevant models and the
results, along with the experimental curves, are shown in Fig. \ref{fig:mz}
.

\section*{A6. Extended modelling}

\subsection{Observation of peak dip feature for some more orientations:}

\begin{figure}
(a)\includegraphics[width=0.45\columnwidth]{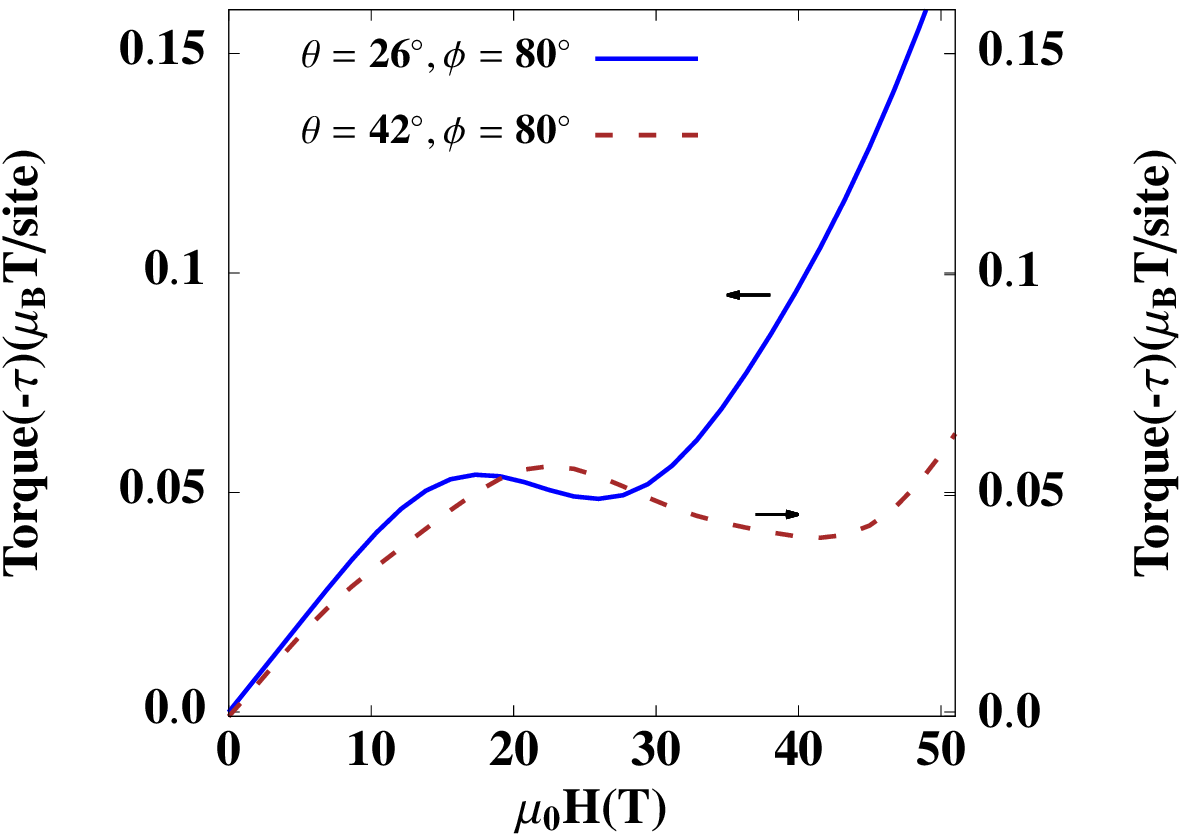}(b)\includegraphics[width=0.45\columnwidth]{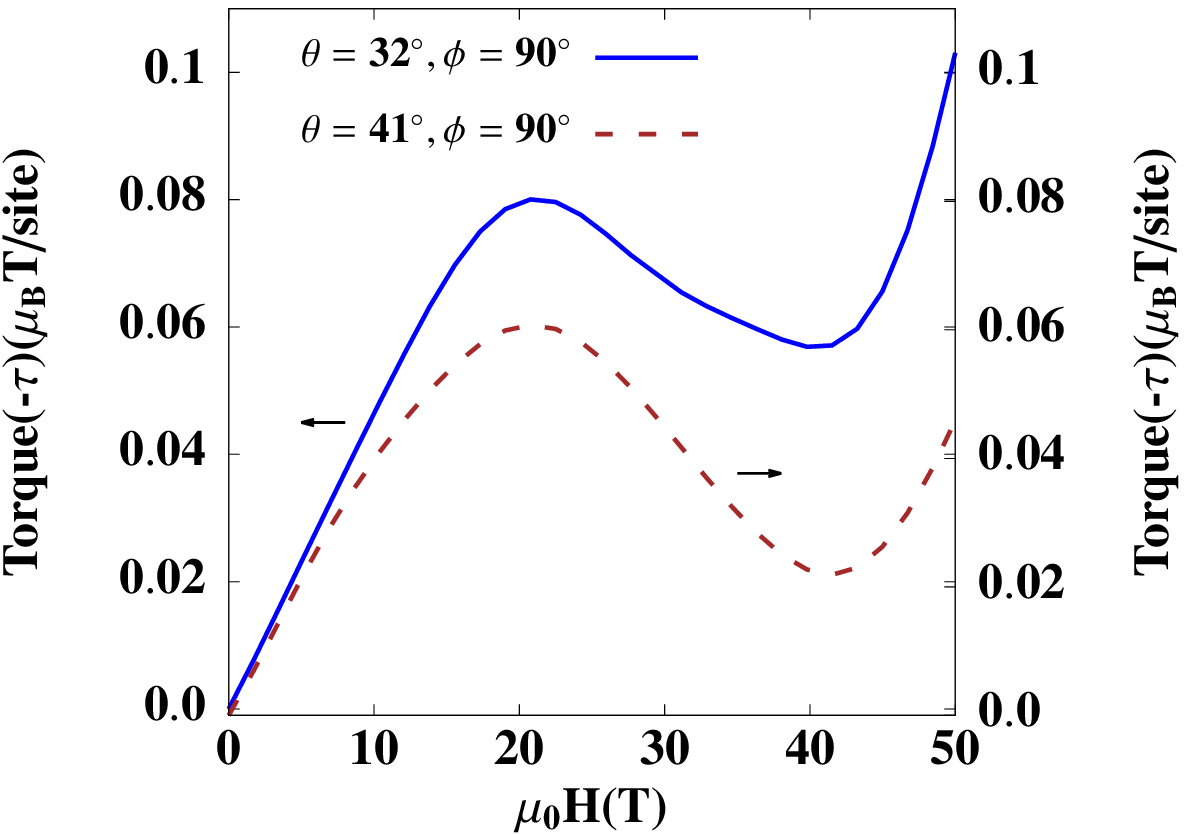}

\caption{\label{fig:modelB}Calculated values of torque for model B with parameters
$J_{h}=2.4$, $J_{k}=-12.0$, $J_{2}=1.6$, $J_{3}=1.2$ (in meV) for
different polar and azimuthal angles as indicated in the figures.
A robust peak-dip feature is observed for a wide range of orientations in model B. (See also the 
contourplot in Fig.6 of the main text.)}
\end{figure}

\begin{figure}
(a)\includegraphics[width=0.45\columnwidth]{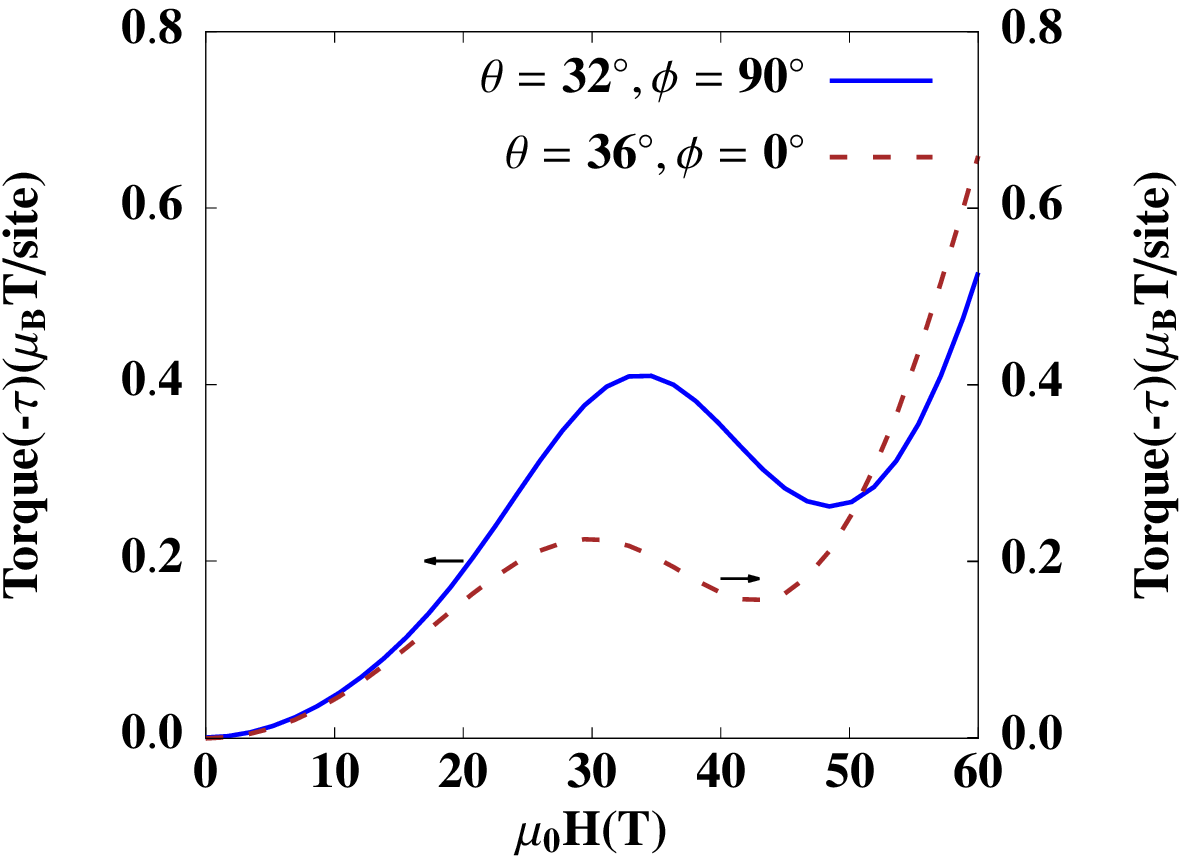}(b)\includegraphics[width=0.45\columnwidth]{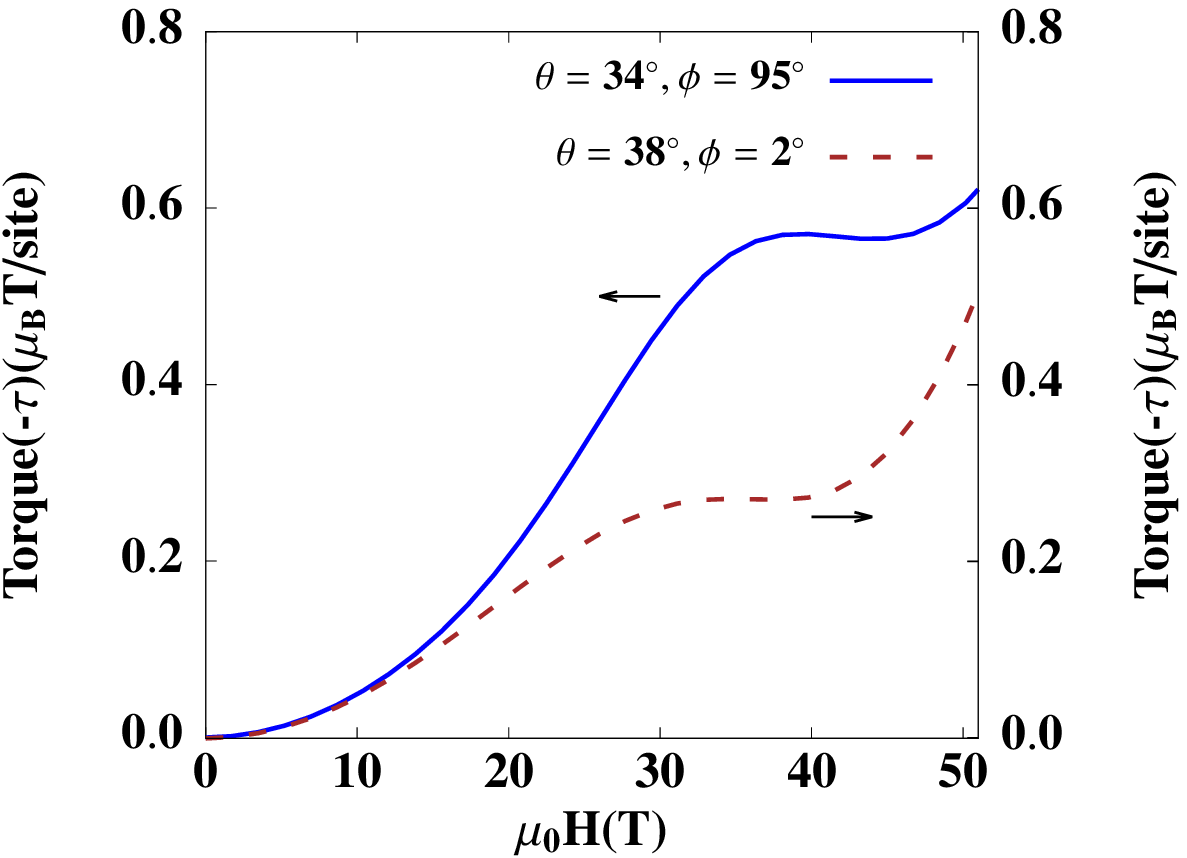}

\caption{\label{fig:model C}Calculated values of torque for model C with parameters
$J_{h}=4.0$, $J_{k}=-16.0$, $\Gamma=2.4$, $\Gamma^{\prime}=-3.2$
(meV) for different polar and azimuthal angles as indicated in the
figure. In (b) the peak-dip feature is present but is shallower than
that observed in (a). }
\end{figure}

Here we consider models B and C of the main text and show the existence
of the peak-dip feature in the torque response for different combinations of polar and azimuthal
angles. This is illustated in figures \ref{fig:modelB} and \ref{fig:model C} for models B and C
respectively. 

\subsection{General absence of a peak-dip feature in models with an antiferromagnetic
Kitaev interaction ($J_{K}>0$):}

\begin{figure}
(a)\includegraphics[width=0.5\columnwidth]{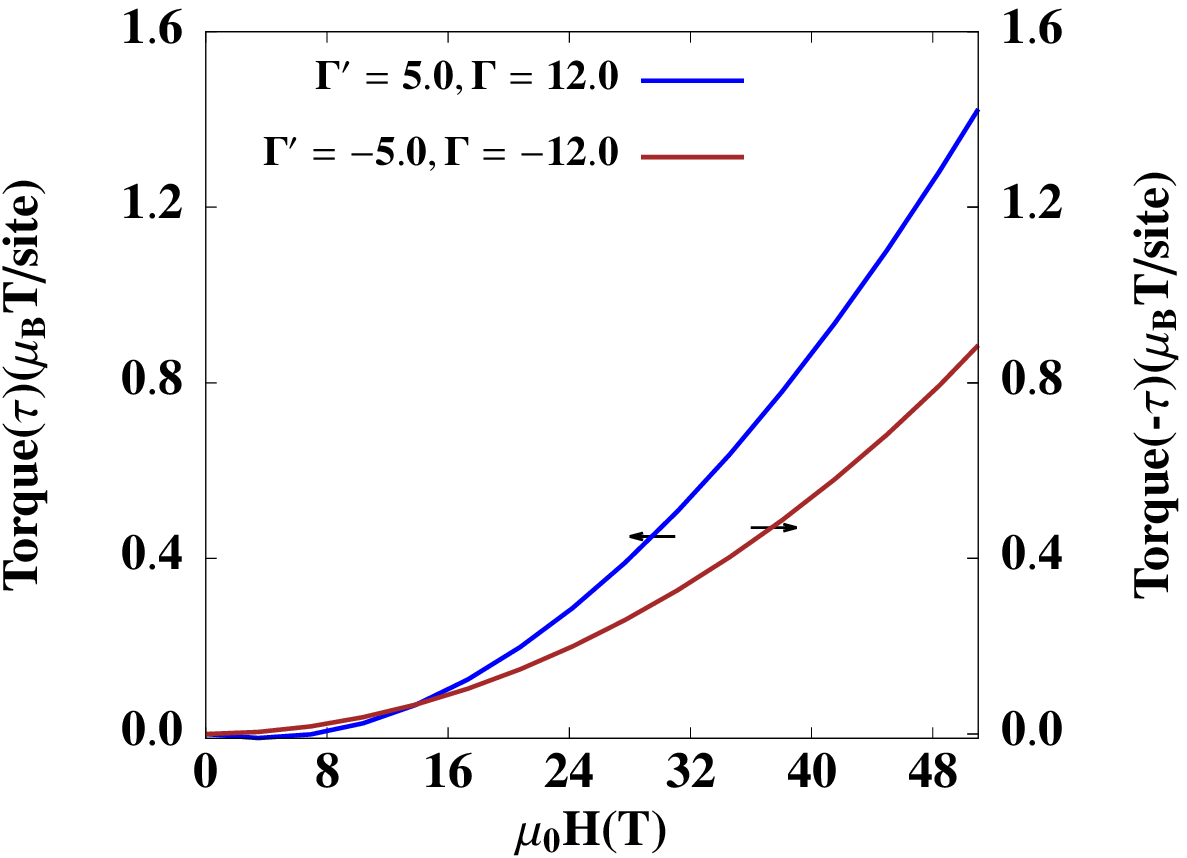}(b)\includegraphics[width=0.5\columnwidth]{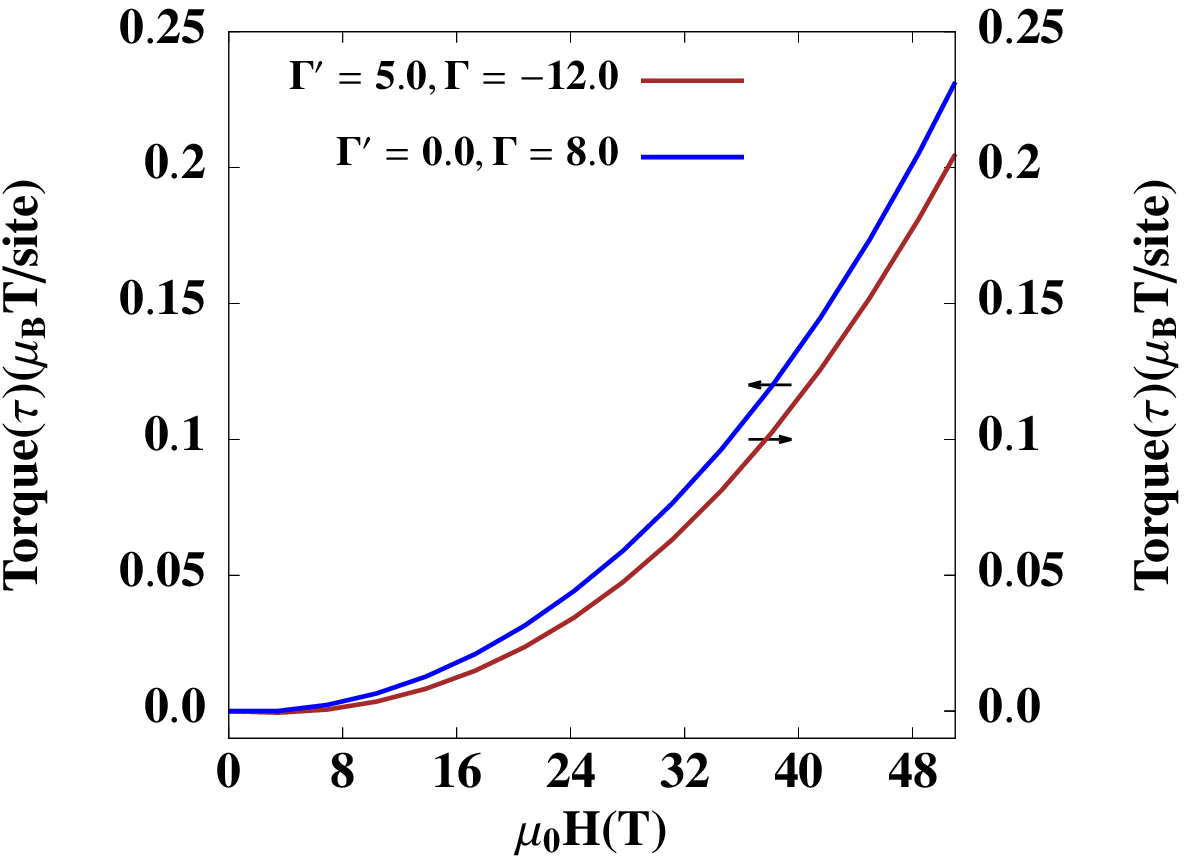}

\caption{\label{fig:jpjr}Calculated values of the torque for models with $J_{h}=-8.0$,
$J_{k}=40.0$(meV), for the orientation $\theta=36^{\circ}$, $\phi=0^{\circ}$,
with $\Gamma$ and $\Gamma^{\prime}$ values as indicated in the figures.
We observe that additional $\Gamma$ and $\Gamma^{\prime}$ terms
do not give rise to any peak-dip features in the torque response. }
\end{figure}

\begin{figure}
(a)\includegraphics[width=0.5\columnwidth]{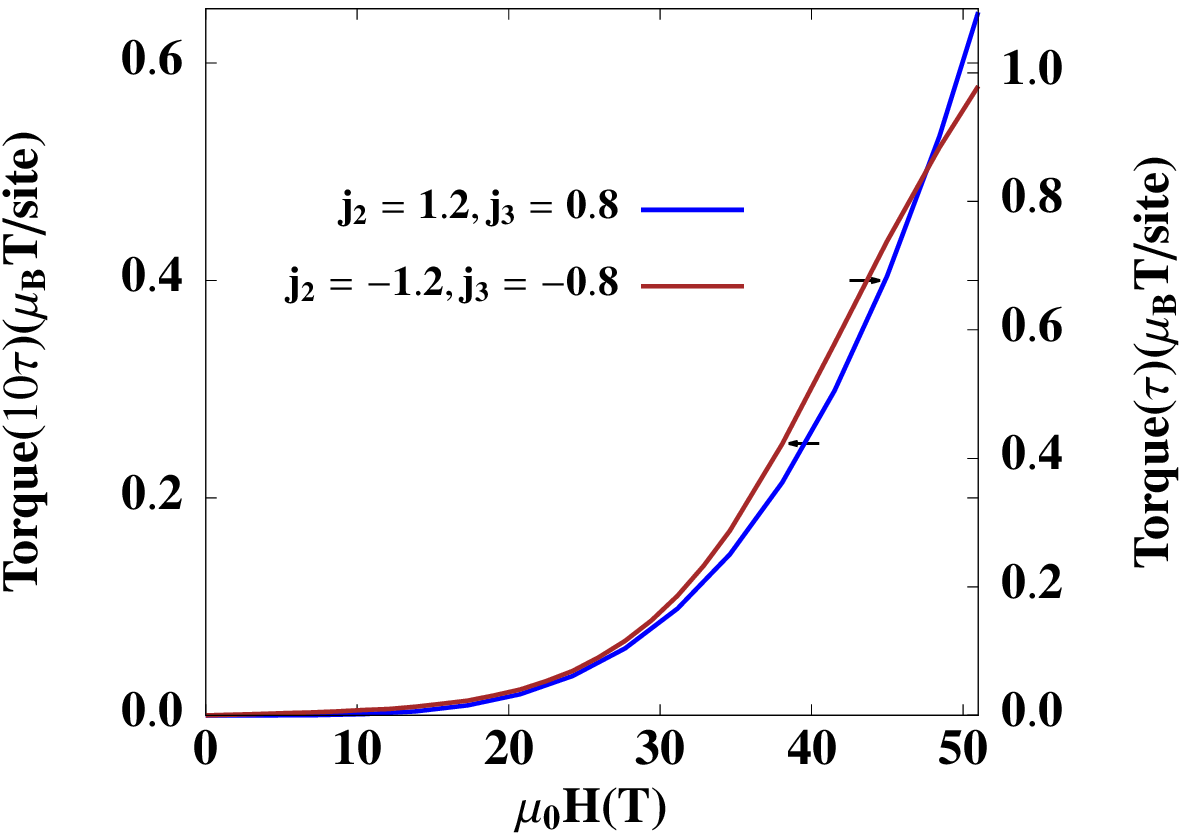}(b)\includegraphics[width=0.5\columnwidth]{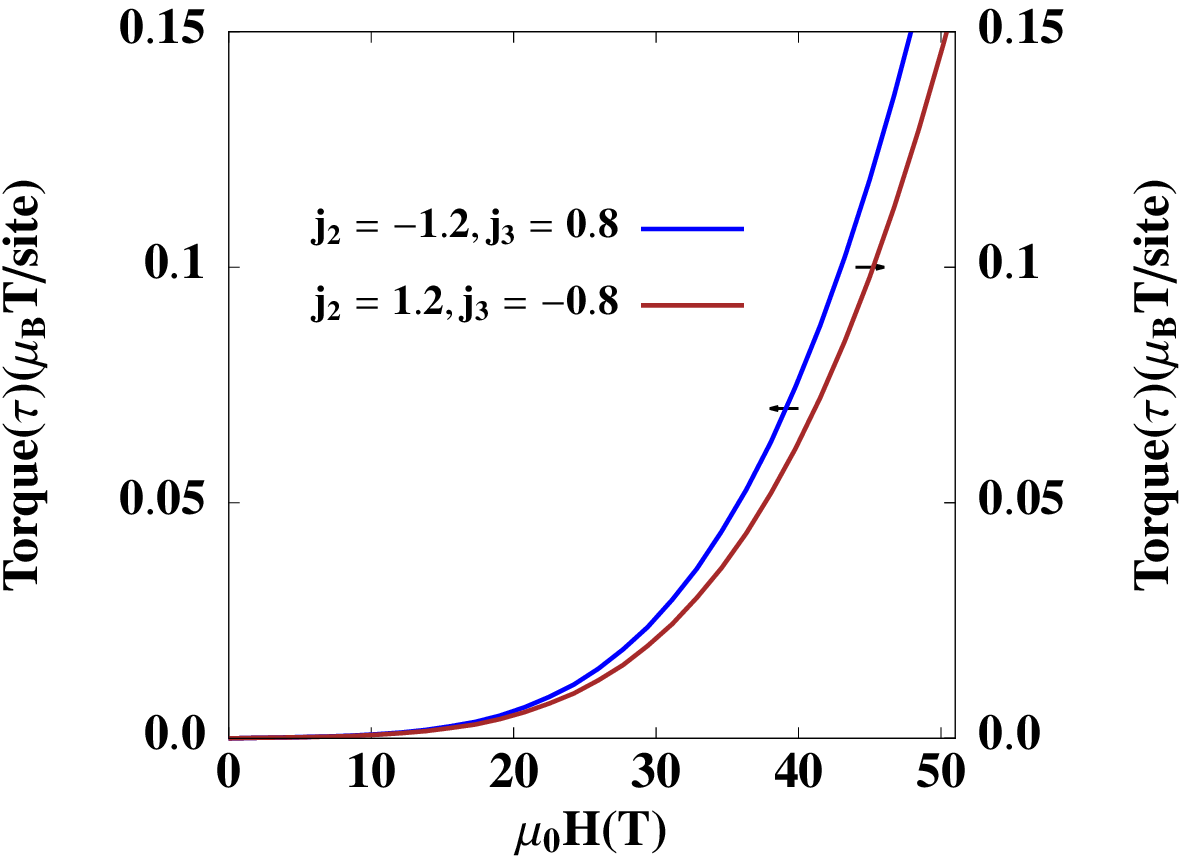}

\caption{\label{fig:j2j3}Here we demonstrate the calculated torque response
for different models with a ferromagnetic Heisenberg and antiferromagnetic
Kitaev interaction with further neighbor Heisenberg interactions.
(a) and (b) correspond to $J_{h}=-4.0$, $J_{k}=21.0$ (meV) for the
orientation $\theta=48^{\circ}$, $\phi=90^{\circ}$, for $J_{2}$
and $J_{3}$ interactions as indicated in the figure. We observe that
further neighbour interactions $J_{2}$ and $J_{3}$ do not give rise
to any peak-dip features in the torque response. }
\end{figure}

\begin{figure}
(a)\includegraphics[width=0.5\columnwidth]{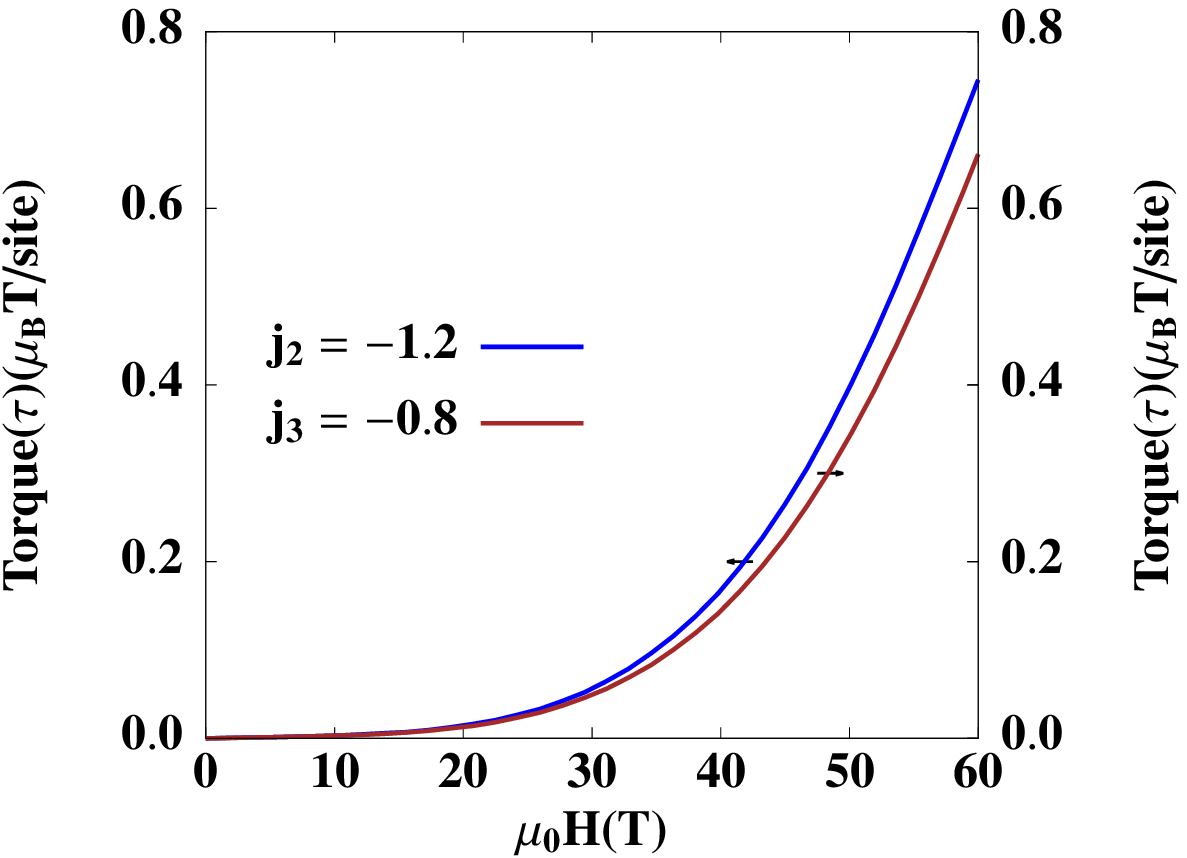}(b)\includegraphics[width=0.5\columnwidth]{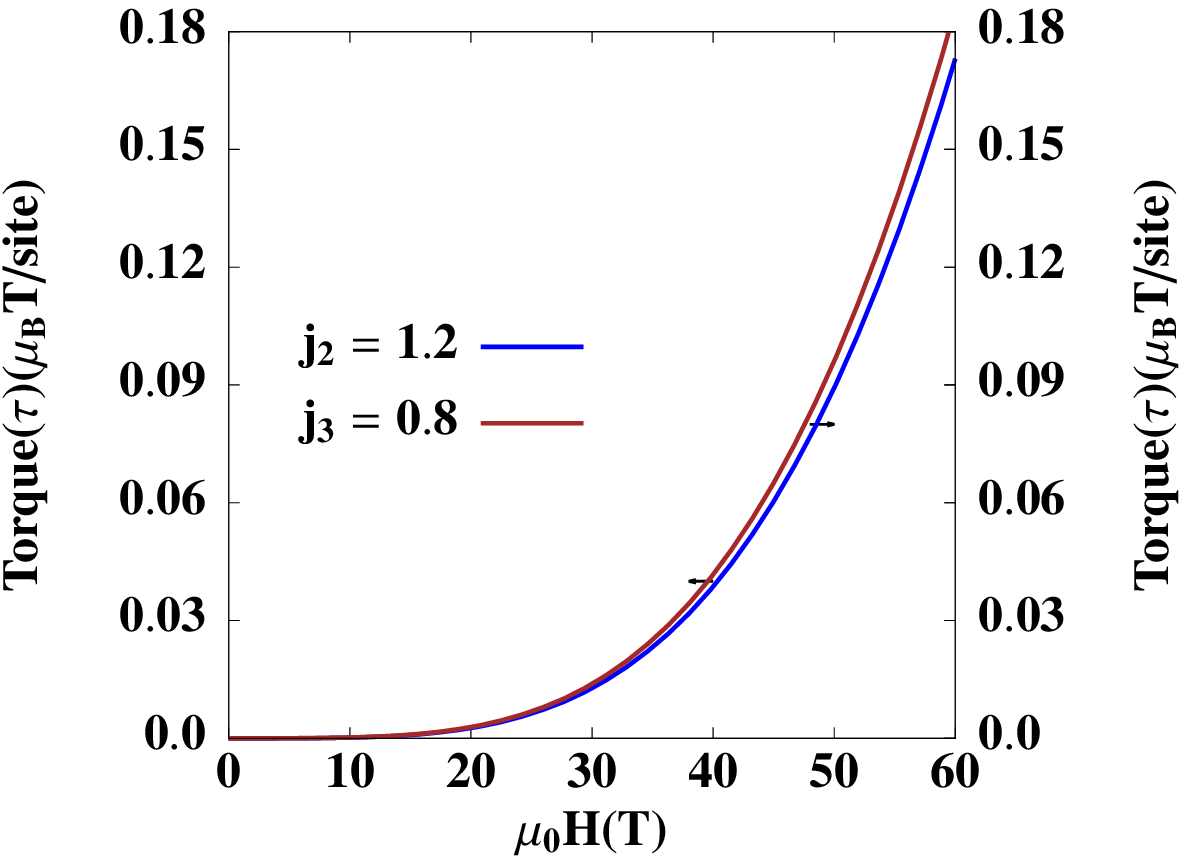}

\caption{\label{fig:j2orj3}Here, we have considered combinations of parameters for $J_{h}=-4.0$,
$J_{K}=21.0$ (in meV) with either $J_{2}$ or $J_{3}$ terms present
but not both. In (a), we consider $J_{2}$ or $J_{3}$ which is ferromagnetic
and in (b), we consider these interactions to be antiferromagnetic, both for the
orientation $\theta=48^{\circ}$, $\phi=90^{\circ}$.
In neither case do we see any peak-dip features in the torque response. }

\end{figure}

\begin{figure}
(a)\includegraphics[width=0.5\columnwidth]{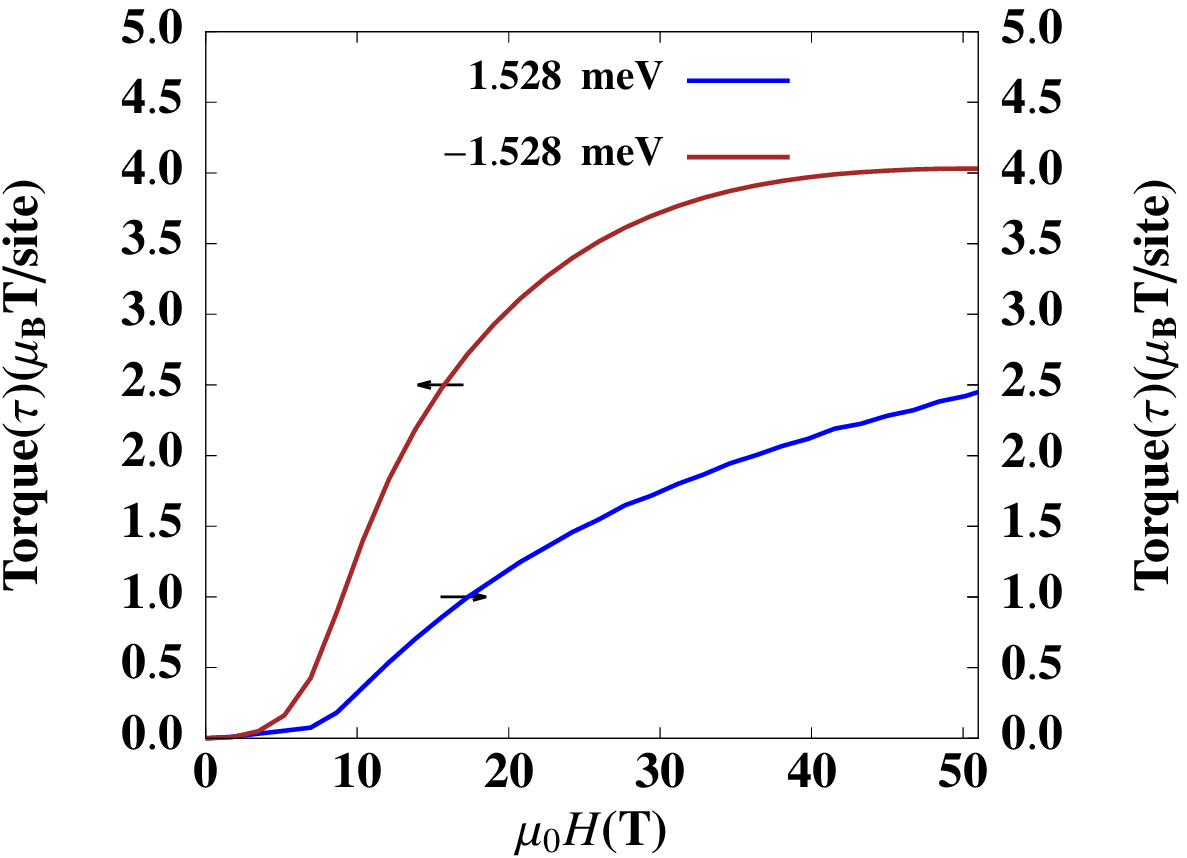}(b)\includegraphics[width=0.5\columnwidth]{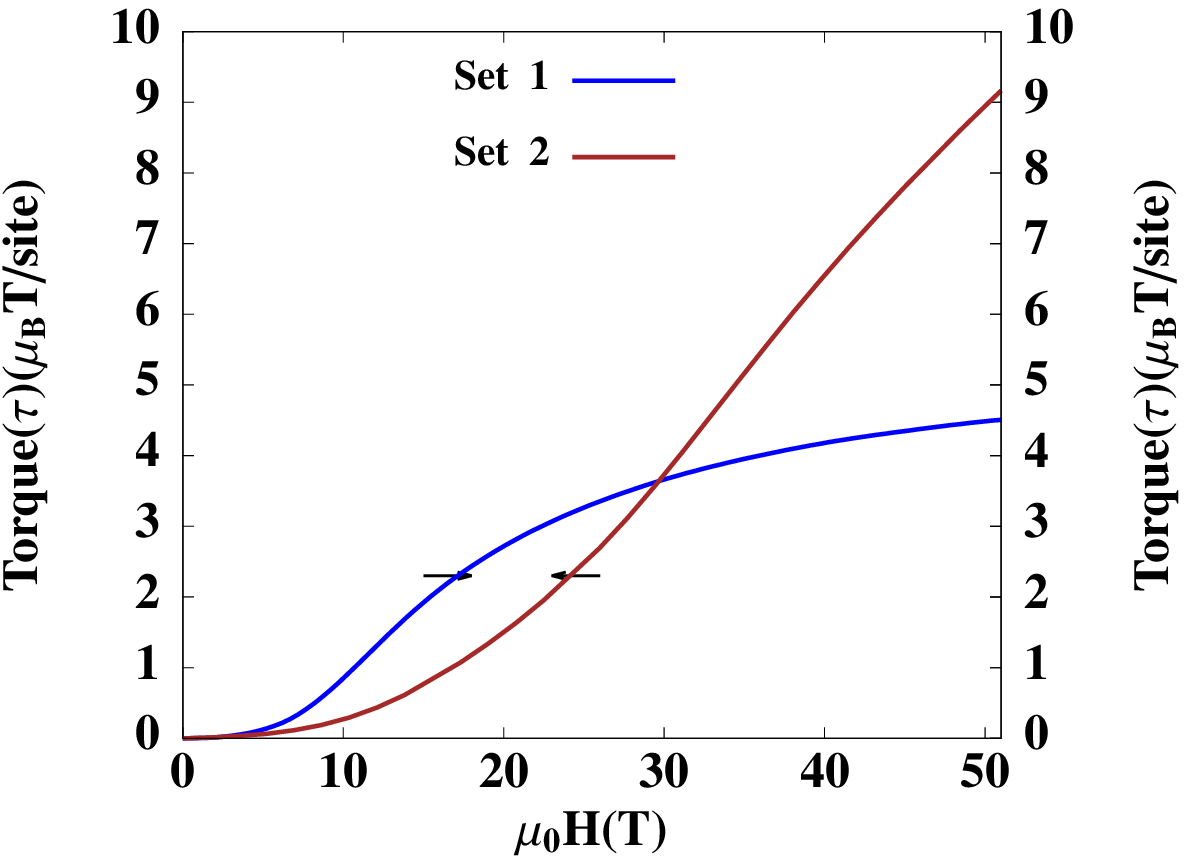}

\caption{\label{fig:gamma}Here we demonstrate the calculated torque response
for different models with a ferromagnetic Heisenberg and antiferromagnetic
Kitaev interaction with additional anisotropic parameters.(a) corresponds
to $J_{h}=-1.84$, $J_{K}=3.2$ (meV) for the orientation $\theta=69^{\circ}$,
$\phi=90^{\circ}$ with an additional $\Gamma$ term taking values indicated
in the figure. (b) shows the torque response for two sets of parameters
with $J_{h}<0$, $J_{K}>0$ and $\Gamma>0$ at different orientations
of the field. Here Set 1 corresponds to $J_{h}=-1.84$, $J_{K}=3.2$
and $\Gamma=1.528$ (meV) for $\theta=36^{\circ}$ and $\phi=0^{\circ}$,
and Set 2 corresponds to $J_{h}=-12.0$, $J_{K}=17.0$, and $\Gamma=12.0$
(meV) for $\theta=48^{\circ}$ and $\phi=90^{\circ}$. }
\end{figure}

The purpose of this section is to show that models with an antiferromagnetic
sign of the Kitaev coupling tuned to a zigzag ground state by a variety
of subleading interactions are generally unable to produce the peak-dip
feature in the torque that is observed in experiment. Figures \ref{fig:jpjr} and \ref{fig:gamma}
illustrate this for models with additional $\Gamma$ and $\Gamma^{\prime}$
interactions, and figures \ref{fig:j2j3} and \ref{fig:j2orj3} for models with various
combinations of antiferromagnetic as well as ferromagnetic further
neighbour interactions $J_{2}$ and $J_{3}$. The different combinations
of parameters considered is summarized in Table I. 

\begin{figure}
(a)\includegraphics[width=0.5\columnwidth]{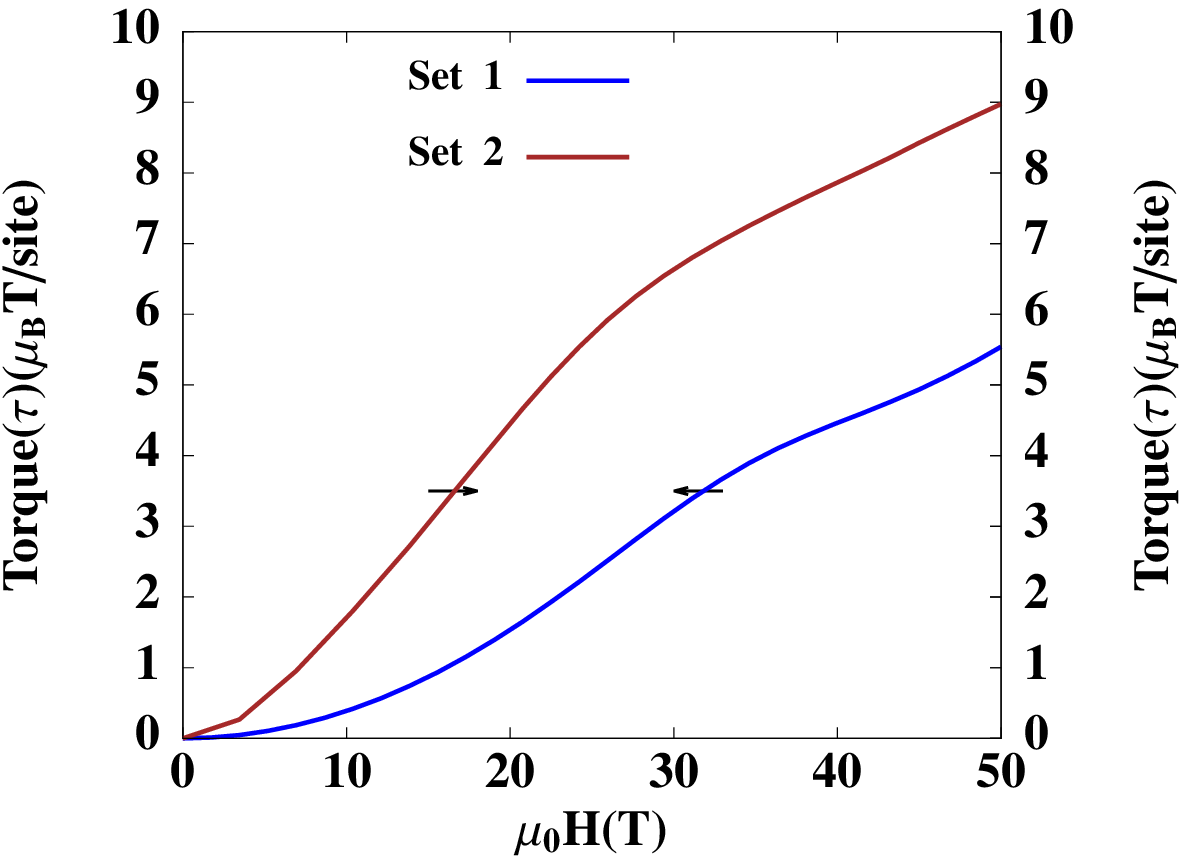}(b)\includegraphics[width=0.5\columnwidth]{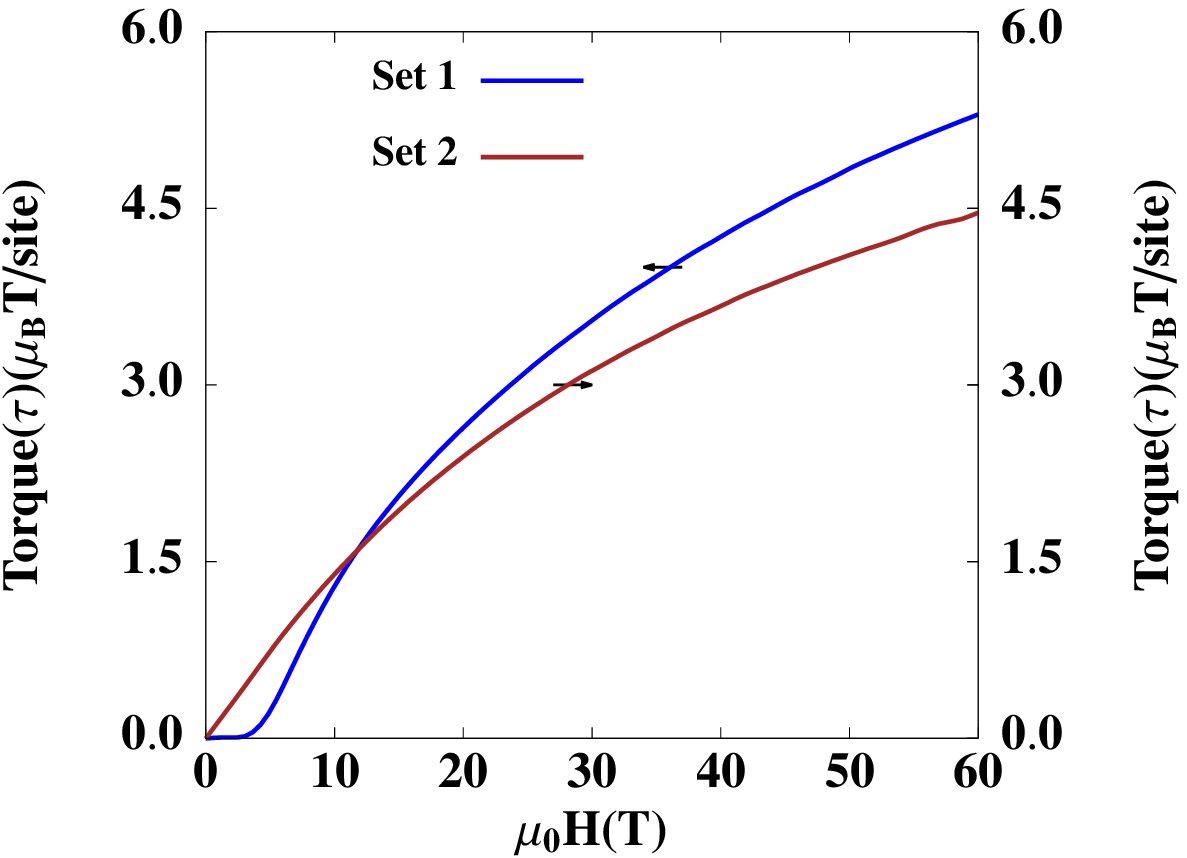}

\caption{\label{fig:fmjh}The figure (a) shows the calculated torque response
for models with both the Kitaev and Heisenberg interactions ferromagnetic.
Here Set 1 corresponds to $J_{h}=-1.0$, $J_{k}=-8.0$ and $\Gamma=4.0$
(in meV) for $\theta=48^{\circ}$ and $\phi=90^{\circ}$ while Set
2 corresponds to $J_{h}=-1.7$, $J_{K}=-6.6$, $J_{3}=2.7$ and $\Gamma=6.6$
(in meV) for $\theta=18^{\circ}$ and $\phi=90^{\circ}$. The figure
(b) shows the calculated torque response for models with $|\Gamma|>|J_{K}|$
with an antiferromagnetic Kitaev and ferromagnetic Heisenberg interaction. Here, Set 1 corresponds
to $J_{h}=-0.98$,$J_{K}=1.17$ and $\Gamma=3.69$ (in meV) while
Set 2 corresponds to $J_{h}=-1.99$, $J_{K}=1.99$ and $\Gamma=2.83$
(in meV), both for $\theta=69^{\circ}$ and $\phi=90^{\circ}$. Clearly, none of the parameter sets with an antiferromagnetic
Kitaev interaction give rise to any peak-dip features in the torque
response. }
\end{figure}

\begin{figure}
(a)\includegraphics[width=0.5\columnwidth]{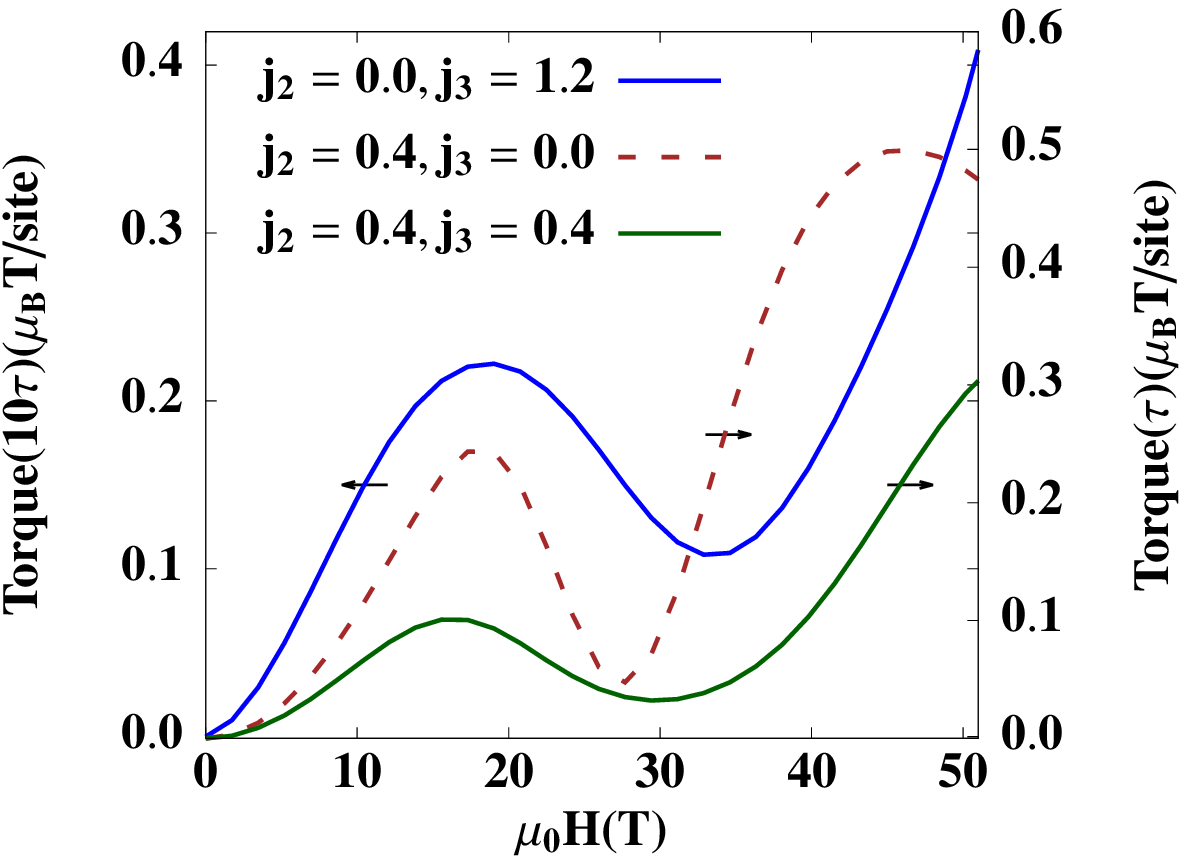}(b)\includegraphics[width=0.5\columnwidth]{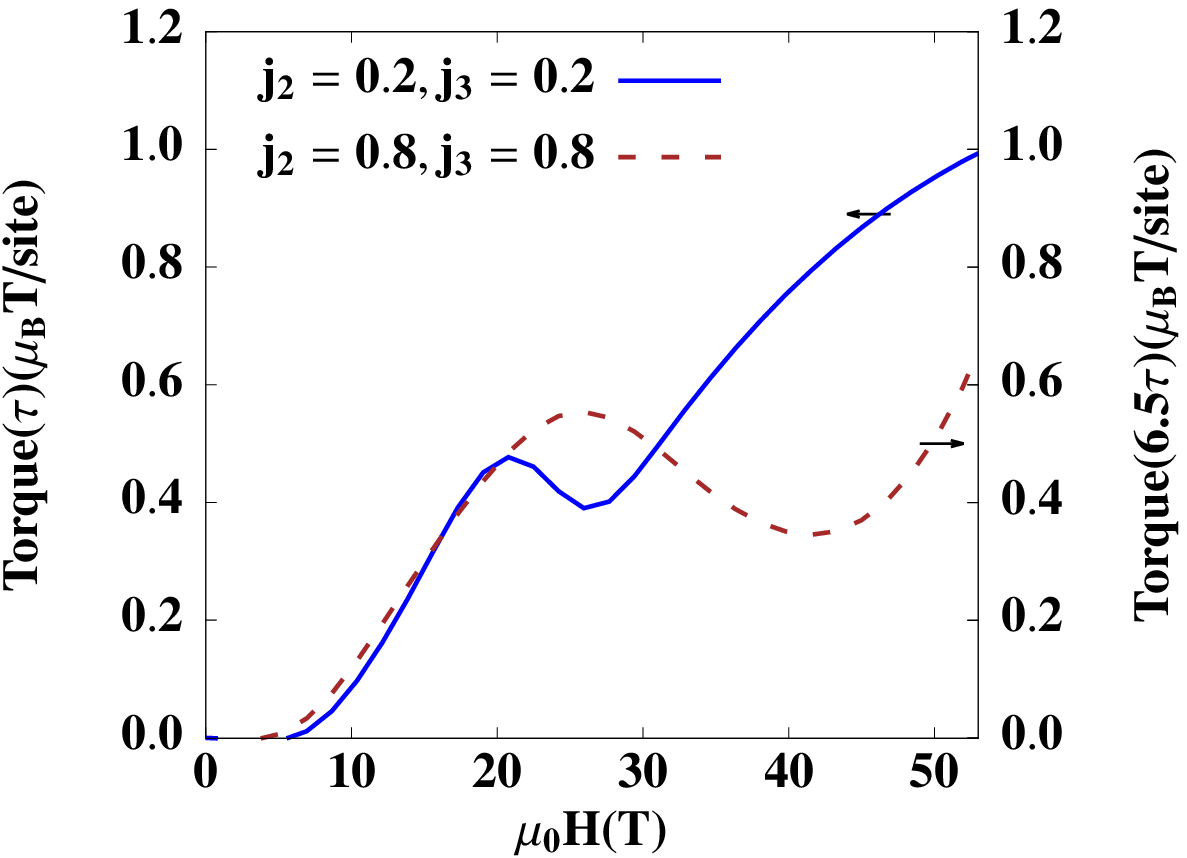}

\caption{\label{fig:nozigzag}The calculated torque response for models with
$J_{h}>0$, $J_{K}<0$ and various combinations of additional interactions
$J_{2}$ and $J_{3}$. Here (a) shows the torque response for $J_{h}=2.4$,
$J_{k}=-20.0$ (meV) for $\theta=36^{\circ}$ and $\phi=0^{\circ}$
with $J_{2}$ and $J_{3}$ values as indicated(in meV), and (b) shows
the torque response for two sets of data where Set 1 corresponds to
$J_{h}=2.4$, $J_{K}=-12.0$, $J_{2}=0.2$ and $J_{3}=0.2$ (in meV)
for $\theta=32^{\circ}$ and $\phi=90^{\circ}$, while Set 2 corresponds
to $J_{h}=2.4$, $J_{K}=-12.0$, $J_{2}=0.8$ and $J_{3}=0.8$ (in
meV) for $\theta=41^{\circ}$ and $\phi=90^{\circ}$. We observe that
peak-dip features, sometimes more than one, are observed in the expected
field range for all of these models, even though most of them do not
exhibit a zigzag ordered ground state. Moreover, the shift in the position
of the peak-dip feature with an increase in the values of the parameters
$J_{2}$ and $J_{3}$ is clearly seen in (b).}
\end{figure}

\begin{figure}
\includegraphics[width=0.9\columnwidth]{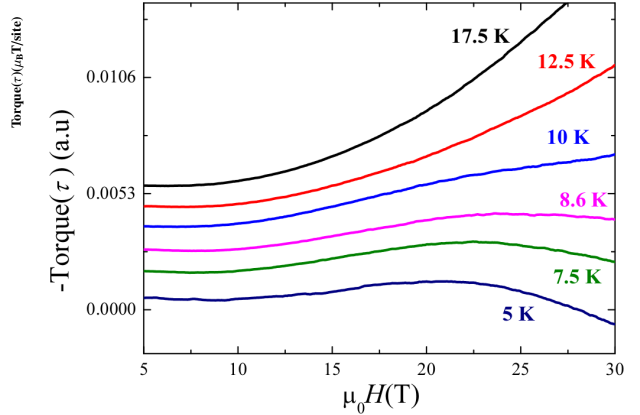}

\caption{\label{fig:temp}The figure shows the torque response corresponding to steady field measurements for $\theta=-20^{\circ}$ and $\phi=0^{\circ}$ at various temperatures in the range
5 K-17.5 K. A shallow peak-dip feature, close to around 20 T, is observed at low temperatures and no longer discernible at temperatures beyond about 12.5 K. This shows that the peak-dip feature is 
associated with a transition from the zigzag ordered ground state to a state with a significantly different torque response.}
\end{figure}

\subsection{Peak-dip features in models with a ferromagnetic Kitaev interaction
($J_{K}<0$):}

Here we consider the torque response for models with a ferromagnetic
Kitaev interaction where the nearest neighbour Heisenberg interaction
is also ferromagnetic, with an additional anisotropic $\Gamma$ and/or
isotropic $J_{3}$ interaction. Such models have been proposed in
the literature for the related Kitaev material $\alpha-$RuCl$_{3}$
which also has a zigzag ground state. We did not see a peak-dip feature
in such models; however, there is a slight flattening of the torque
response curve at intermediate fields. This is illustrated in Fig.
\ref{fig:fmjh}(a).

\subsection{Peak-dip features in the absence of zigzag order:}

Here we demonstrate that models with a ferromagnetic Kitaev $J_{K}$,
antiferromagnetic Heisenberg $J_{h}$ and additional antiferromagnetic
further-neighbour interactions $J_{2}$ and $J_{3}$ can give rise
to peak-dip features even in the absence of zigzag order in the
ground state. The presence of the peak-dip features thus provides an
independent handle which can distinguish the response of such models
from those with an antiferromagnetic Kitaev interaction. This is illustrated
in Fig. \ref{fig:nozigzag}, where we find that the peak-dip feature
is observed even for those combinations of parameters where either
$J_{2}$ or $J_{3}$ vanishes, or $J_{2}$, $J_{3}$ are both small
as compared to the nearest-neighbour interactions $J_{h}$ and $J_{K}$.
Such combinations of parameters often do not give rise to a zigzag
ordered ground state, and cannot be used to represent Na$_{2}$IrO$_{3}$,
but they still do give rise to prominent peak-dip features in the
torque. Table \ref{tab:tab2} summarizes the different models we have
considered, with a ferromagnetic Kitaev interaction and additional
subdominant terms. 

\begin{ruledtabular}
\begin{table}
\begin{tabular}{|c|c|c|c|c|c|}
\hline 
Model A variant & $\Gamma$ & $\Gamma^{\prime}$ & $J_{2}$ & $J_{3}$ & Ref.(if any)\tabularnewline
\hline 
\hline 
1. With $\Gamma,\Gamma^{\prime}$ & + & - & $\times$ & $\times$ & \cite{rau2014trigonal}\tabularnewline
\hline 
 & - & + & $\times$ & $\times$ & \cite{rau2014trigonal}\tabularnewline
\hline 
 & - & - & $\times$ & $\times$ & \cite{rau2014trigonal}\tabularnewline
\hline 
2. With $J_{2},J_{3}$ & $\times$ & $\times$ & + & + & \tabularnewline
\hline 
 & $\times$ & $\times$ & - & - & \tabularnewline
\hline 
 & $\times$ & $\times$ & + & - & \tabularnewline
\hline 
 & $\times$ & $\times$ & - & + & \tabularnewline
\hline 
 & $\times$ & $\times$ & - & $\times$ & \tabularnewline
\hline 
 & $\times$ & $\times$ & $\times$ & - & \tabularnewline
\hline 
 & $\times$ & $\times$ & + & $\times$ & \tabularnewline
\hline 
 & $\times$ & $\times$ & $\times$ & + & \tabularnewline
\hline 
3. With $\Gamma$ & + & $\times$ & $\times$ & $\times$ & \cite{janssen2017magnetization},\cite{rau2014generic}\tabularnewline
\hline 
 & - & $\times$ & $\times$ & $\times$ & \cite{janssen2017magnetization},\cite{rau2014generic}\tabularnewline
\hline 
\end{tabular}

\caption{\label{tab:tab1}For model A ($J_{h}<0$, $J_{K}>0$), introduction
of various additional terms $\Gamma$,$\Gamma^{\prime}$, $J_{2}$
and $J_{3}$, with a zigzag ground state, does not reveal any peak-dip
features. The + and - indicate the sign of the coupling and $\times$
denotes the absence of the corresponding coupling. Thus, for a wide
variety of parameters, antiferromagnetic Kitaev couplings do not produce
the observed peak-dip feature. }
\end{table}
\end{ruledtabular}

\begin{ruledtabular}
\begin{table}
\begin{tabular}{|c|c|c|c|c|c|c|}
\hline 
 & $J_{h}$ & $J_{2}$ & $J_{3}$ & $\Gamma$ & $\Gamma^{\prime}$ & Peak-dip(present/absent)\tabularnewline
\hline 
\hline 
Model B & + & + & + & $\times$ & $\times$ & Yes\tabularnewline
\hline 
 & + & + & $\times$ & $\times$ & $\times$ & Yes\tabularnewline
\hline 
 & + & $\times$ & + & $\times$ & $\times$ & Yes\tabularnewline
\hline 
Model C & + & $\times$ & $\times$ & + & - & Yes\tabularnewline
\hline 
Ref.\cite{janssen2017magnetization} & - & $\times$ & + & + & $\times$ & No\tabularnewline
\hline 
 & - & $\times$ & $\times$ & + & $\times$ & No\tabularnewline
\hline 
\end{tabular}

\caption{\label{tab:tab2}For models with $J_{K}<0$ (model B, model C as well
as other models with various anisotropic and Heisenberg interactions),with
a zigzag ground state, the peak-dip feature may or may not be present.
The + and - indicate the sign of the coupling and $\times$ denotes
the absence of the corresponding coupling. The last two sets of parameters
have been suggested in Ref. \cite{janssen2017magnetization} and correspond
to a ferromagnetic Heisenberg interaction. The remaining ones correspond
to an antiferromagnetic Heisenberg interaction. Clearly, the presence
of zigzag order in models with $J_{K}<0$ does not necessarily produce
the peak-dip feature. Thus, the peak-dip feature is an independent
tool to constrain the parameter space of possible effective Hamiltonians. }
\end{table}

\end{ruledtabular}

\subsection{Evolution of the peak-dip feature as a function of polar angle $\theta$
and field $H$:}

Here we discuss the evolution of the torque response for model B as
a function of the polar angle $\theta$ and field value $H$, by presenting
a contourplot of the first derivative of the torque, $\frac{d\tau}{dH}$
(for the torque $\tau$ and field $H$), and compare our results with
the experimental data in Fig.2 of the main text. We find that for
model B, a peak-dip feature is robustly observed for all orientations
$\theta$ for a given value of the azimuthal angle $\phi$, and the
position as well as the shape of the peak-dip evolves as a function
of $\theta$, as expected from the experiment. Theoretically, the transverse magnetization (torque)
response could well be negative, and in such cases, we plot $-\frac{d\tau}{dH}$
instead, as we are not interested in the absolute value of the torque
obtained, but only in the position of the peak-dip, which is indicated
by the regions where the first derivative of the torque changes sign.
Fig.6 in the main text illustrates our results, and it is clear that
although there is qualitative agreement with the experimental results,
unlike the actual data, the distance between the peak and the dip, i.e. the width of the region of
nonmonotonicity increases at extreme values of $\theta$.

\section*{A7: Temperature-dependence of the torque response}

The torque response was measured capacitively for different temperatures in the range 5 K-17.5 K, at steady fields up to 30 T for the orientation $\theta=-20^{\circ}$, $\phi=0^{\circ}$.
A peak-dip feature observed close to about 20 T at low temperatures becomes indiscernible beyond a temperature of about 12.5 K. This is illustrated in Fig.
\ref{fig:temp}. The exact detection of the temperature at which this feature disappears is limited by the resolution of the measurement, as well 
as temperature resolution very close to the zigzag ordering temperature. Our results do however establish that the peak-dip feature is present only below the
zigzag ordering temperature and is therefore related to a transition from the long-range ordered ground state.


\end{document}